\documentclass[aps,pra,superscriptaddress,
nofootinbib,
twocolumn,
preprintnumbers,
floatfix]{revtex4}

\usepackage{amsmath,amsfonts,amsbsy,amssymb,amsthm}
\usepackage{mathrsfs,bbm}
\usepackage{natbib}
\usepackage{mathtools}
\usepackage[pdftex]{graphicx}
\usepackage[pdftex,colorlinks]{hyperref}
\usepackage[caption=false]{subfig}

\newcommand{\ket}[1]{\lvert #1 \rangle}
\newcommand{\bra}[1]{\langle #1 \rvert}
\newcommand{\ketbra}[2]{\ket{#1}\bra{#2}}
\newcommand{\expt}[1]{\langle #1 \rangle}

\newcommand{\smallfrac}[2]{\mbox{$\frac{#1}{#2}$}}
\newcommand{\dg}{^\dagger}

\newcommand{\op}[2]{\ket{#1}\bra{#2}}

\newcommand{\half}{\tfrac{1}{2}}

\newcommand{\erf}[1]{Eq.~(\ref{#1})}

\begin{document}

\preprint{YITP-21-136}

\title{ 
Modified coherence of quantum spins in a damped pure-dephasing model}

\author{ Mattias T. Johnsson }	
\affiliation{Department of Physics and Astronomy, Macquarie University, North Ryde, NSW 2109, Australia}

\author{ Ben Q. Baragiola }
\affiliation{Yukawa Institute for Theoretical Physics, Kyoto University, Kitashirakawa Oiwakecho, Sakyo-ku, Kyoto 606-8502, Japan}
\affiliation{Centre for Quantum Computation and Communication Technology, School of Science, RMIT University, Melbourne, VIC 3001, Australia}
\affiliation{Centre for Engineered Quantum Systems, Department of Physics and Astronomy, Macquarie University, North Ryde, NSW 2109, Australia}

\author{Thomas Volz}
\affiliation{Department of Physics and Astronomy, Macquarie University, North Ryde, NSW 2109, Australia}
\affiliation{Centre for Engineered Quantum Systems, Department of Physics and Astronomy, Macquarie University, North Ryde, NSW 2109, Australia}

\author{Gavin K. Brennen}
\affiliation{Department of Physics and Astronomy, Macquarie University, North Ryde, NSW 2109, Australia}
\affiliation{Centre for Engineered Quantum Systems, Department of Physics and Astronomy, Macquarie University, North Ryde, NSW 2109, Australia}

\begin{abstract}
We consider a spin-$j$ particle coupled to a structured bath of bosonic modes that decay into thermal baths.
We obtain an analytic expression for the reduced spin state and use it to investigate non-Markovian spin dynamics. In the heavily overdamped regime, spin coherences are preserved due to a quantum Zeno affect. 
We extend the solution to two spins and include coupling between the modes,  which can be leveraged for preservation of the symmetric spin subspace.
For many spins, we find that inter-mode coupling gives rise to a privileged symmetric mode gapped from the other modes. This provides a handle to selectively address that privileged mode for quantum control of the collective spin. 
Finally, we show that our solution applies to defects in solid-state systems, such as NV$^{-}$ centres in diamond.

\end{abstract}

\maketitle

\section{Introduction}

Quantum control of spins is now an advanced field with applications being developed for quantum sensing \cite{RevModPhys.89.035002} and quantum computing \cite{doi:10.1063/5.0007444}. Less well-developed, however, is the control of the environments with which the spins inevitably interact. In the limit where a spin is only weakly coupled to its environment, which has a large bandwidth relative to the spin dynamics, then the Born-Markov and rotating-wave approximations apply. The environment quickly loses information, and the Markovian system dynamics obey a Lindblad master equation~\cite{Gardiner1985, BreuerBook}. 
In another setting, the coupling is weak but the environmental correlations are long lived. Dynamical decoupling pulses can be employed to protect the spins~\cite{PhysRevA.58.2733}.
These limits are starting points for approaches to studying reduced spin dynamics for potential engineering of spin control and coherence preservation.
More generally, though, one must consider that both (a) the spin-environment coupling is not weak and (b) the environment is ``structured" in that it possesses non-trivial temporal correlations and modified spectral density \cite{Rey2013}.

The fundamental tools for studying such systems beyond weak coupling are bipartite spin-boson models, where the spin is strongly coupled to an environment of modes~\cite{Grifoni1998, Thorwart2004}. 
Various techniques have been employed to study reduced-spin dynamics including generalized master equations~\cite{Grifoni1998}, hierarchical methods~\cite{Yoshitaka2005, Hughes2012, Strunz2019}, dilation to a tripartite unitary dynamics~\cite{PhysRevLett.120.030402}, and others~\cite{Wilhelm2004, Loss2005, Lovett2018, Nori2019}.
A subset of spin-boson models are pure dephasing models (also called the independent boson models), where the spin experiences no energy exchange with the modes. 
This arises in a variety of physical settings including exciton-phonon dynamics~\cite{Gilmore2005, Pouthier2013} and defects in crystal lattices~\cite{BetzTorrBien14}.
Pure dephasing models admit exact solutions for the spin dynamics~\cite{MahanBook,MukamelBook, Chaudhry2013}, revealing non-Markovian dephasing that strongly depends on the modes' spectral density and initial state~\cite{Gan_2010, Breuer2012}.

In this work, we consider a large-spin ($j>1/2$) pure-dephasing spin-boson model with an additional feature: the modes themselves decay irreversibly into thermal baths. 
We present an exact analytic solution for the reduced spin in this setting, variations of which are plentiful in the literature~\cite{Chaudhry2013,Pouthier2013,Chen2010} but do not combine both large spin and mode dissipation.
The effects of mode dissipation can be pronounced --- in the overdamped regime, they significantly enhance spin-coherence lifetimes in analogy to quantum-Zeno-type effects when measurements are performed~\cite{Chaudhry2018}.

For multiple spins, the spectrum and decay rates of the bosonic modes can induce effective interactions between the spins. We consider a
multi-spin setting motivated by defects in solid-state system, where each electronic emitter couples dominantly to local vibrational modes~\cite{Gali_2011, AlbrRetzJele13, BetzTorrBien14, karim2020}. 
Coupling between local modes (indicative of non-local normal modes) can induce a separation of energy scales that implies distinct dynamics on different collective subspaces of the spins. Similar effects are found in spin-boson studies of exciton dynamics using effective modes~\cite{Burghardt2009}, where the existence of a single or group of ``preferred'' modes can lead to lengthened electronic coherences~\cite{Michele2014}. 
Selectively addressing preferred modes provides a handle for quantum control of collective spin degrees of freedom~\cite{Chaudhry2013}. 

The paper is organized as follows. In Sec.~\ref{secAnalyticDephasingResults} we introduce the model and solve for the exact reduced dynamics of a single large spins, and we illustrate important limiting behaviour including an overdamped setting that preserves spin coherences using a quantum-Zeno-type effect. 
In Sec.~\ref{sec:DissipativeProtectionOfCoherences} we solve for the reduced dynamics of two spins in the same setting where the bosonic modes are themselves coupled to each other. We analyze how the symmetric spin subspace can be preserved for a longer time due to this modification of the environment. Extending the analysis to many spins, we show in Sec.~\ref{sec:structured} that coupling between all the bosonic modes can in certain regimes open a gap between a symmmetric eigenmode and the rest, and this provides a mechanism for coherent control on the symmetric spin subspace. 
In Sec.~\ref{sec:NVExample}, we show that our solution is also useful to describe the physics of solid-state defects, such as NV$^{-}$ centres in diamond. We include two other relevant effects in such systems: pure dephasing and (optical) decay of the spin. Finally, we conclude with a summary of results and suggestions for further applications.

\section{Large spin coupled to a collection of vibrational modes: analytic solution}
\label{secAnalyticDephasingResults}

Our starting point is a closed system comprising a single spin-$j$ particle coupled to a collection of harmonic oscillators. Although the results apply in general for spin-boson coupling, we consider for concreteness the harmonic oscillators to be a discrete set of vibrational modes determined by the boundary conditions in a crystal setting.
We derive an analytic formula for the time-evolved joint state of the closed system from which we extract the reduced state of the spin by tracing over the modes. Since the interaction with the vibrational modes is unitary (and not dissipative), the reduced spin state experiences non-Markovian effects. 

 Spin dephasing arises from state-dependent coupling to the set of local vibrational modes, because the electronic excited state deforms the local electron density of the crystal. The Hamiltonian for this situation is the (large-spin) spin-boson model,
    \begin{equation}
        \hat{H} = \Omega \hat{j}_z 
        + \sum _k \omega_k ( \hat{v}^{\dagger}_k \hat{v}_k + \tfrac{1}{2} )
        + \hat{j}_z \otimes \sum _k (\eta_k \hat{v}^{\dagger}_k + \eta^* _k\hat{v}_k)
\label{eqSpinMutipleHOModeCoupling1}
    \end{equation}
where $\eta_k$ characterizes the interaction strength between the spin and the $k$th vibrational mode and $\hbar = 1$.

The spin is described by a set of $2j+1$ bare eigenstates satisfying
    \begin{equation} \label{spineigenbasis}
        \hat{j}_z \ket{j,m} = m \ket{j,m}
    \end{equation}
with transition frequency $\Omega$.
Each vibrational mode is described by bosonic field operators satisfying $[\hat{v}_k, \hat{v}^\dagger_{k'}] = \delta_{k,k'}$.

The joint state of the spin-boson system at time $t$ is formally given by
\begin{equation} \label{jointstate}
    \hat \rho(t) = \hat{U}(t) \hat \rho_{\mathrm{spin}}(0) \otimes \hat \rho_V(0) \hat{U}^{\dagger} (t) ,
\end{equation}
where $\hat \rho_\mathrm{spin}(0) \otimes \hat \rho_V(0)$ is the initial joint state.
The interaction-picture propagator for the spin-boson system (with respect to the free Hamiltonians of the spin and vibrational modes) is
    \begin{equation}
        \label{originalpropagator}
        \hat{U}(t) = {\cal{T}} \exp \left[ -i \int _0 ^t dt'   \hat{j}_z \otimes \hat{V}(t') \right] ,
    \end{equation}
where ${\cal{T}}$ designates the time-ordering operator, and the Hermitian interaction-picture mode operator is
\begin{equation}
    \hat{V}(t) \coloneqq \sum _k (\eta_k \hat{v}^{\dagger}_k e^{i \omega_k t} + \eta_k^*\hat{v}_k e^{-i \omega_k t}).
\end{equation}

By writing the propagator in the eigenbasis of the spin, we can manipulate it into a form that is useful for calculating time evolution: 
    \begin{align} \label{timeorderedpropagator}
        \hat{U}(t) 
                   = \sum_{m=-j}^j {\cal{T}} \exp \left[ -i m \int _0 ^t dt' \,  \hat{V}(t') \right] \op{j,m}{j,m} .
    \end{align}
The time-ordered integral in this expression can be simplified. Using the fact that all the vibrational mode operators for $k \neq k'$ commute, we remove the time-ordering by employing a Magnus expansion, which terminates at second order.
Details are given in Appendix~\ref{appendix:Magnus}. This gives the expression, 
\begin{align}
        {\cal{T}}  \exp & \left[ -i m \int_0^t dt'  \, \hat{V}(t') \right] = e^{ i m^2 \Phi(t)}  \exp \left[ -i m \int_0^t dt'\, \hat{V}(t') \right]  ,
\label{propagatorv2}
\end{align}
where the $c$-number phase is 
    \begin{equation} \label{cnumberphase}
        \Phi(t)  
    \coloneqq - \int_0^t \! dt_1 \int_0^{t_1} dt_2 \int d\omega \, J(\omega) \, \sin [\omega (t_1 -t_2)]
    \end{equation} 
and we have defined a spectral density,
    \begin{equation} \label{spectraldensity}
        J(\omega) \coloneqq \sum_k |\eta_k|^2 \, \delta(\omega - \omega_k).
    \end{equation}
Importantly, the time ordering has been removed, and the phase factor is determined only by the spectral density (through the coupling strengths in the Hamiltonian) and not by the state of the vibrational modes.


With the propagator in Eq.~(\ref{propagatorv2}), we can express the general solution for the joint spin-vibrational state at time $t$ as Eq.~\eqref{jointstate}:
    \begin{align}
    \label{jointstatesolution}
       \hat \rho(t) 
       &= \sum_{m,m'=-j}^{j}e^{ i (m^2 - m'^2)\Phi(t)} \rho_\text{spin}^{m,m'}(0)  
        \op{j,m}{j,m'} \nonumber \\
       &\quad \quad \quad \otimes e^{ -i m \int_0^t dt'\, \hat{V}(t')} \hat \rho_V(0) e^{i m' \int_0^t dt'\, \hat{V}(t')}  ,
    \end{align}
where the matrix elements of the initial spin state are
    \begin{equation}
        \rho_\text{spin}^{m,m'}(0) \coloneqq \bra{j, m} \hat \rho_\text{spin}(0) \ket{j,m'}
         ,
    \end{equation}  
and we have used that $\hat{V}^\dagger(t) = \hat{V}(t)$, since it's a Hamiltonian.


\subsection{Reduced state of the spin}
The reduced density matrix for the spin at time $t$, $\hat \rho_{\mathrm{spin}}(t)$, is found by tracing over the vibrational degrees of freedom in the expression for the joint state, Eq.~\eqref{jointstate},
    \begin{equation} \label{reducedspinstate_def}
        \hat \rho_{\mathrm{spin}}(t) = \mathrm{Tr}_V \left[ \hat{\rho}(t) \right]  .
    \end{equation}
By decomposing the reduced spin state in the eigenbasis, Eq.~\eqref{spineigenbasis},
    \begin{equation}
       \hat{\rho}_\text{spin}(t) = \sum_{m,m'=-j}^{j} \rho_\text{spin}^{m,m'}(t) \op{j,m}{j,m'}  ,
    \end{equation}    
we find the matrix elements by tracing over the vibrational modes in the general solution, Eq.~\eqref{jointstatesolution},
    \begin{align} \label{spinmatrixelementstimet}
        \rho^{m,m'}_\text{spin}(t) = \rho^{m,m'}_\text{spin}(0) e^{ i (m^2 - {m'}^2) \Phi(t)} \mathcal{S}(t) .
    \end{align}
For each matrix element, labeled by $m$ and $m'$, this expression requires evaluating the term
    \begin{align} \label{traceintegral}
            \mathcal{S}(t) \coloneqq {\mathrm{Tr}}_V \Big[ e^{ -i (m-m') \int_0^t dt'\, \hat{V}(t')}  \hat \rho_V(0) \Big]   ,
    \end{align}
for a given initial state of the vibrational modes $\hat{\rho}_V(0)$. 

We consider the situation where the vibrational modes are initially in a thermal state characterized by $\beta = 1/k_B T$ for temperature $T$. The initial state across the $k$ modes is given by
   \begin{equation} \label{thermalstatetensorprod}
        \hat \rho_V(0) = \bigotimes_k \hat \rho_{\text{therm},k},
    \end{equation}
where the thermal state for vibrational mode $k$ is given in the diagonal coherent-state basis ($P$-function) as
    \begin{align} \label{thermalstatek}
        \hat \rho_{\text{therm},k} 
         = \frac{1}{\pi \bar{n}_k} \int d^2\alpha \, \exp \left( -\frac{|\alpha|^2}{\bar{n}_k} \right) \op{\alpha}{\alpha} ,
    \end{align} 
with thermal occupation $\bar{n}_k = [\exp( \beta \omega_k) - 1]^{-1}$.
In this case, the integral in Eq.~\eqref{traceintegral} can be evaluated analytically. Details following the method of Agarwal~\cite{Agarwal2010} are given in Appendix \ref{Appendix:thermaltrace}. 
Plugging the result, Eq.~\eqref{appendixSfinal}, into the general formula, Eq.~\eqref{spinmatrixelementstimet}, gives the matrix elements of the reduced spin state, 
\begin{align}
\rho_{\mathrm{spin}}^{m \, m'} (t) &= \rho_{\mathrm{spin}}^{m \, m'}(0) \exp  \Bigg[ - \int_0^t dt_1 \int_0^{t_1} dt_2 \int_0^\infty d\omega \, J(\omega) \nonumber \\
&\times \Big( i (m^2-m'^2)  \sin [\omega (t_1 -t_2)]  
 \nonumber \\
& \quad + (m-m')^2 \coth \big( \tfrac{ \beta   \omega}{2} \big) \, \cos[ \omega (t_1 -t_2)] \Big) \Bigg].
\label{eqrhoMMprimeInitial}
\end{align}
These dynamics are nontrivial for the spin coherences, while the diagonal matrix elements ($m = m'$) do not evolve. 
For small bath temperatures, $\beta \rightarrow \infty$, this equation approaches the situation where the spin coherences dynamically evolve along with the vibrational modes, but they do not experience decay and revival. 
Finally, we note that, although we have focused on the reduced spin state $\hat{\rho}(t)$ the dynamical map above can be applied to any operator by decomposing it in the $\hat{j}_z$-basis.

We can also consider Eq.~(\ref{eqrhoMMprimeInitial}) as arising from the correlation functions of the mode operators. 
Defining a quadrature operator for mode $k$ (giving the unnormalized position quadrature when $\eta_k$ is real),
\begin{equation} \label{quadratureX}
    \hat{X}_k \coloneqq \eta_k \hat v_k + \eta^*_k \hat v_k^\dagger ,
\end{equation}
we note that if one has a Hamiltonian of the form in \erf{eqSpinMutipleHOModeCoupling1}, and the initial state of the vibrational modes is thermal, then the quadrature correlation function of the modes
is given by (see Appendix~\ref{Appendix:correlation}),
\begin{align} \label{thermalcorrelationfunction}
    C(t) &\coloneqq \sum_k \expt{\hat{X}_k(t) \hat{X}_k(0)} \\
     &= \sum_k |\eta_k|^2 \big[ \coth \big( \tfrac{\beta \omega_k} {2} \big) \cos ( \omega_k t) - i \sin (\omega_k t) \big] \\
    &= \int_0^\infty d\omega J(\omega) \left[ \coth \big( \tfrac{\beta  \omega}{2} \big) \cos(\omega t) - i \sin(\omega t) \right] .
\end{align}
In the final line, we have expressed the correlation function in terms of the spectral density $J(\omega)$, \erf{spectraldensity}.
It will also convenient to divide the correlation function into its real and imaginary parts
    \begin{align} \label{correlationfunc_realimag}
        C(t) = C_{\mathrm{Re}}(t) + i C_{\mathrm{Im}}(t).
    \end{align}
Using the above expressions, the analytic form for the reduced-spin matrix elements, Eq.~\eqref{eqrhoMMprimeInitial}, can also be written as

    \begin{align}
        \rho^{\mathrm{spin}} _{m \, m'} (t) &= \rho^{\mathrm{spin}} _{m \, m'}(0)  \exp \big[ i (m^2-m'^2) {\cal{I}}_{\mathrm{Im}}(t; \vec \omega) \nonumber \\
        &  - (m-m')^2 {\cal{I}}_{\mathrm{Re}}(t; \vec \omega) \big]  ,
    \label{eqSingleSpinsmmprime1}
    \end{align}
where we have defined integrals over the real and imaginary parts of the correlation function,
    \begin{subequations} \label{fancyintegrals}
    \begin{align}
        {\cal{I}}_{\mathrm{Re}}(t; \vec \omega) &\coloneqq  \int_0^t \! dt_1 \! \int_0^{t_1} \! dt_2 \, C_\text{Re}(t_1-t_2), \\
        {\cal{I}}_{\mathrm{Im}}(t; \vec \omega) &\coloneqq \int_0^t \! dt_1 \! \int_0^{t_1} \! dt_2 \, C_\text{Im}(t_1 - t_2) ,
    \end{align}    
    \end{subequations}    
with $\vec{\omega}$ included to indicate that each integral is a function of the mode frequencies. Recall that this solution is in the interaction picture with respect to the bare spin and bare mode Hamiltonians.

The imaginary part $\mathcal{I}_\text{Im}$ gives the coherent dynamics of the spin coherences, and the real part $\mathcal{I}_\text{Re}$ describes their decay.
Note that the integrals in \erf{fancyintegrals} can in principle be evaluated term-by-term by recognizing that the correlation function, \erf{thermalcorrelationfunction}, is a sum over the vibrational mode index $k$. That is, we may express the integrals as
\begin{subequations} \label{integralsassums}
\begin{align}
    {\cal{I}}_{\mathrm{Re}}(t; \vec{\omega}) &= \sum_k {\cal{I}}_{\mathrm{Re}}(t; \omega_k), \\
        {\cal{I}}_{\mathrm{Im}}(t; \vec{\omega}) &= \sum_k {\cal{I}}_{\mathrm{Re}}(t; \omega_k) .
\end{align}
\end{subequations}
This form will be valuable for evaluating the terms below.

\subsection{Spin dephasing in the presence of thermal dissipation of the vibrational modes} \label{thermaldissipationsection}

Above, we considered a single large spin and a collection of vibrational modes evolving unitarily as a closed system. Here, we generalize this situation to an open system where each vibrational mode is coupled to a local dissipative bath at inverse temperature $\beta_k$.
This is described by the master equation for joint state $\hat{\rho}$,
    \begin{eqnarray}
    \frac{d}{dt} \hat{\rho} &=&  -\frac{i}{\hbar} \big[ \hat{H}, \hat{\rho} \big] + \sum_k  \mathcal{D}^{\text{th}}_k[\hat{\rho} ] ,
    \end{eqnarray}
with spin-boson Hamiltonian in Eq.~\eqref{eqSpinMutipleHOModeCoupling1} and thermal dissipator
    \begin{align}
     \mathcal{D}^{\text{th}}_k[\hat{\rho}] &\coloneqq 
    \Gamma_k  (\bar{n}_k+1) \, (\hat{v}_k \rho \hat{v}_k^{\dagger} 
    - \tfrac{1}{2}\hat{v}_k^{\dagger} \hat{v}_k \rho  - \tfrac{1}{2}\rho \hat{v}_k^{\dagger} \hat{v}_k ) \nonumber
    \\
    & \quad + \Gamma_k  \bar{n}_k \, ( \hat{v}_k^{\dagger} \rho \hat{v}_k 
    - \tfrac{1}{2} \hat{v}_k \hat{v}_k^{\dagger} \rho - \tfrac{1}{2} \rho \hat{v}_k \hat{v}_k^{\dagger} )  .
    \label{dissipator}
    \end{align}
The top line describes loss of vibrational excitations into the bath, and the second line describes incoherent heating according to the temperature of the bath (note that this term vanishes when the bath occupancy vanishes.) 
Going to the interaction picture with respect to the free Hamiltonian of the spin and of the vibrational modes does not affect the form of the thermal dissipator. 

In the previous section we demonstrated that the evolution of the reduced spin density matrix can be described by the thermal-state correlation functions of the modes.
This is true even when the vibrational modes decay according to the Markovian thermal dissipator in Eq.~\eqref{dissipator}, which gives a decaying correlation function. 
Including the dissipator causes the quadratures to decay via the replacement $\hat{v}_k \rightarrow  e^{-\frac{\Gamma_k t}{2}} \hat{v}_k$ in Eq.~\eqref{quadratureX}.

We now find the correlation function, \erf{thermalcorrelationfunction}, for the vibrational modes. There are in principle two temperatures associated with each vibrational mode: that of the initial vibrational-mode states and that of the bath to which each mode couples. We set these to be the same under the assumption that each vibrational mode is initially in equilibrium with its local bath.
As derived in Appendix~\ref{Appendix:correlation}, the correlation function [\erf{thermalcorrelationfunction}] of the decaying vibrational modes is
\begin{align}
    C(t) 
     &= \sum_k |\eta_k|^2 e^{-\frac{1 }{2}\Gamma_k t } \big[ \coth \big( \tfrac{\beta \omega_k} {2} \big) \cos ( \omega_k t) - i \sin (\omega_k t) \big].
      \label{eqCorrelationFunctionSomoza}
\end{align}

\begin{figure*}[t]
  \includegraphics[width=1.9\columnwidth]{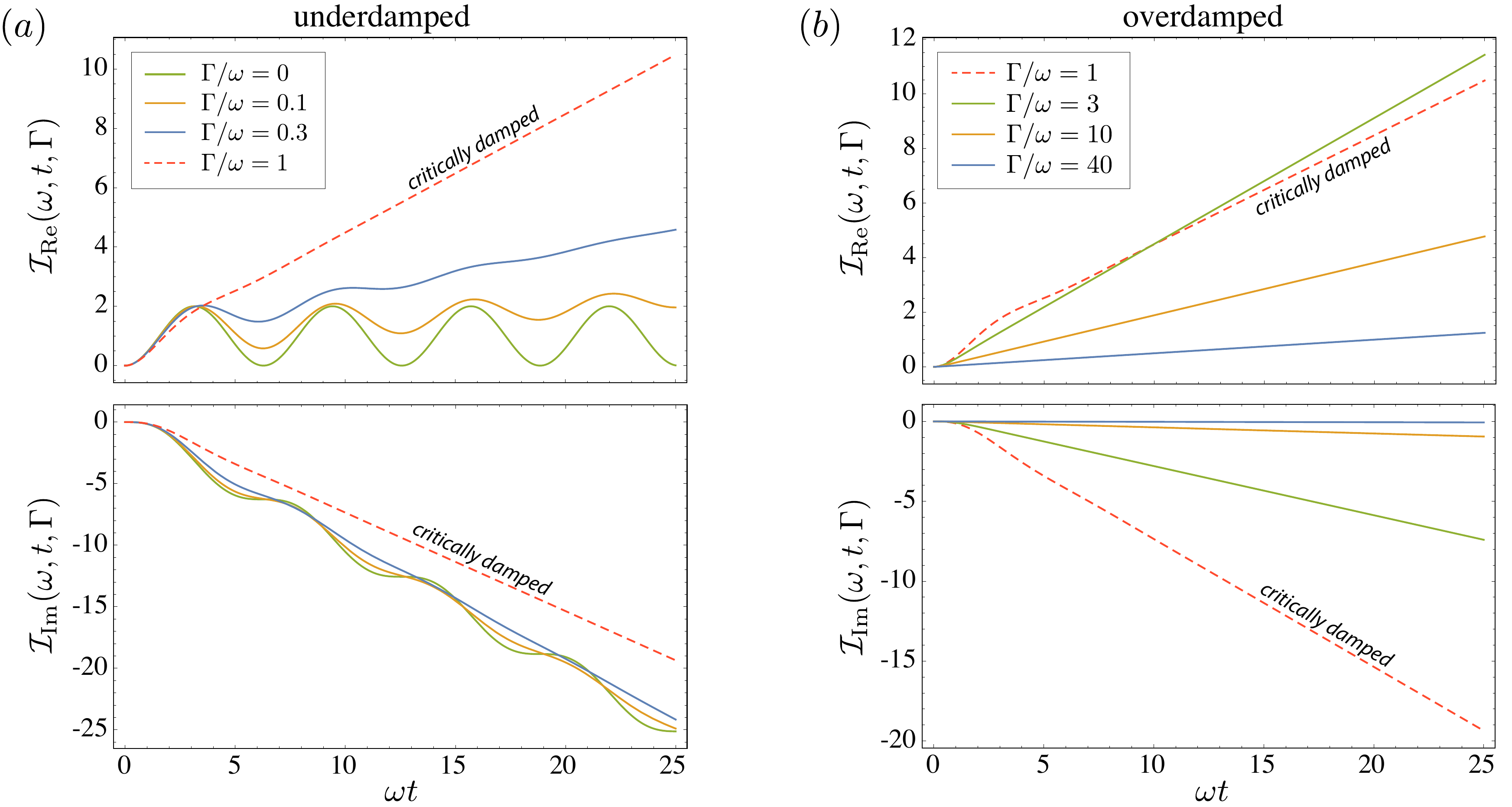}
\caption{
Dephasing factor $\mathcal{I}_\text{Re}(t; \omega, \Gamma)$ and unitary phase factor $\mathcal{I}_\text{Im}(t; \omega, \Gamma)$ in Eq.~\eqref{integralfactors} as a function of time for the (a) underdamped and (b) overdamped regimes. Parameters are $\eta/\omega = 1$, $\beta \rightarrow \infty$ (zero temperature), and critical damping occurrs at $\Gamma/\omega = 1$. The amplitude of a spin coherence between $m$ and $m'$ is determined by  $\mathcal{I}_\text{Re}$ [Eq.~\eqref{eqRhoMMprimeSingleSpin}], which in the underdamped regime increases as $\Gamma/\omega$ does. However, in the overdamped regime, increasing $\Gamma/\omega$ decreases the $\mathcal{I}_\text{Re}$, thus producing less spin dephasing. In this regime, the accumulated unitary phase between $m$ and $m'$, determined by $\mathcal{I}_\text{Im}$ also decreases with $\Gamma/\omega$ --- in the heavily overdamped regime, spin coherences are frozen and experience no dephasing or unitary-type evolution at all.
}
\label{fig:Factors}
\end{figure*}

While the open-systems dynamics of the joint state is Markovian, the spin subsystem evolves in a non-Markovian way. The expression for the reduced spin density matrix is given by \erf{eqSingleSpinsmmprime1},
    \begin{align}
        \rho^{\mathrm{spin}} _{m \, m'} (t) = & \rho^{\mathrm{spin}} _{m \, m'}(0)  \exp \big[ i (m^2-m'^2) \mathcal{I}_{\mathrm{Im}}(t; \vec \omega, \vec \Gamma) \nonumber \\
        &  - (m-m')^2 {\cal{I}}_{\mathrm{Re}}(t; \vec \omega, \vec \Gamma) \big]  ,
    \label{eqRhoMMprimeSingleSpin}
    \end{align}
with the integrals in \erf{fancyintegrals} taken here over the real and imaginary parts of the correlation function in \erf{eqCorrelationFunctionSomoza} that now includes thermal dissipation. Note that we include an additional label on the integrals $\vec{\Gamma}$ to include the vibrational-mode decay rates.
We now evaluate the integrals above term-by-term, as described by \erf{integralsassums}. For vibrational mode $k$ with frequency $\omega_k$ and decay rate $\Gamma_k$, the integrals evaluate to
\begin{widetext}
\begin{subequations} \label{integralfactors}
\begin{align} 
\mathcal{I}_{\mathrm{Re}}(t; \omega_k, \Gamma_k) 
&= 
    \frac{2 |\eta_k|^2 \coth \big(\frac{\beta  \omega_k}{2}\big)} {\left(\Gamma_k ^2+4 \omega_k ^2\right)^2}
   \left\{
    - 2(\Gamma_k ^2 - 4\omega_k ^2) +\Gamma_k t( \Gamma_k^2 +4 \omega_k^2) 
   +e^{-\frac{\Gamma_k  t}{2}}\left[-8 \Gamma_k  \omega_k \sin (\omega_k t)
   + 2 \left(\Gamma_k ^2-4 \omega_k ^2\right) \cos (\omega_k t
   )
   \right] \right\}
 \label{eqIRe} \\ 
{\cal{I}}_{\mathrm{Im}}(t; \omega_k, \Gamma_k) 
&=   
   \frac{4 |\eta_k| ^2 }{\left(\Gamma_k ^2+4 \omega_k ^2\right)^2}
   \left\{ 4 \omega_k \Gamma_k
   - \omega_k t (\Gamma_k ^2 + 4 \omega_k^2 )
   -e^{-\frac{\Gamma_k  t}{2}} \left[ \left(\Gamma_k ^2-4 \omega_k
   ^2\right) \sin (\omega_k t )+4 \Gamma_k 
   \omega_k  \cos (\omega_k t
   )\right] \right\}  .
   \label{eqIIm}
\end{align}
\end{subequations}
\end{widetext}
These expressions, 
which we refer to as the dephasing factor ($\mathcal{I}_\text{Re}$) and the unitary phase factor ($\mathcal{I}_\text{Im}$), 
complete the full non-Markovian description of the reduced spin state, \erf{eqRhoMMprimeSingleSpin}. 
The factors depend on various parameters in the Hamiltonian, but they are independent of the spin size and the coherences between specific spin-basis states labeled by $m$ and $m'$. Rather, the spin size $j$ sets the bounds for $m$ and $m'$, and these enter the dynamical solution as multiplicative factors $(m^2-m'^2)$ and $(m-m')^2$ in Eq.~\eqref{eqRhoMMprimeSingleSpin}. Fig.~\ref{fig:Factors} shows $\mathcal{I}_\text{Re}$ and $\mathcal{I}_\text{Im}$ for a single mode and various mode decay rates. The effects on spin coherences can be organized into underdamped $\Gamma < \omega$ and overdamped $\Gamma \gg \omega$ regimes, as discussed in the caption and further below. Note that for vanishing damping rates, $\Gamma_k \rightarrow 0$, the expressions above simplify to
\begin{align}
    \mathcal{I}_{\mathrm{Re}}(t; \omega_k, 0) 
    &= \frac{2 |\eta_k|^2}{\omega_k^2} \coth \big( \tfrac{ \beta  \omega_k }{2} \big) \sin^2 \big( \tfrac{t \omega_k}{2} \big) \\
    \mathcal{I}_{\mathrm{Im}}(t; \omega_k, 0) 
    &= \frac{|\eta_k|^2}{\omega_k^2} \big[ \sin \big( \tfrac{t \omega_k}{2} \big) - \omega_k t\big] .
    \label{nodecayfactors}
\end{align}

In pure dephasing models, the diagonal matrix elements of the spin do not evolve, see the $m=m'$ terms in \erf{eqRhoMMprimeSingleSpin}. This is due to the fact that the Hamiltonian, \erf{eqSpinMutipleHOModeCoupling1}, is diagonal in $\hat{j}_z$. The off-diagonal spin coherences, in contrast, experience both unitary-type and dephasing-type dynamics according to $\mathcal{I}_{\mathrm{Im}}(t; \vec \omega, \vec \Gamma)$ and $\mathcal{I}_{\mathrm{Re}}(t; \vec \omega, \vec \Gamma)$, respectively --- this is pure dephasing. Notice that the accumulated phase from the unitary-type dynamics is trivial for a two-level spin ($j=\frac{1}{2}$), since $m^2 - m'^2$ always vanishes. Thus, pure spin dephasing is most clearly illustrated as shown in Fig.~\ref{fig:SingleSpinUnderAndOverDamped}.
Given a larger spin, $j>\frac{1}{2}$, the accumulated phases are not trivial. In that case, the unitary-type dynamics are akin to a single-axis twisting Hamiltonian~\cite{ma2011}, which generates spin squeezing for a spin with $j>\frac{1}{2}$. 
For larger spins, these spin squeezing effects are always in competition with the dephasing-type dynamics, since they scale in the same way with the spin-mode coupling $\eta_k$.

\subsubsection*{Asymptotic regimes} \label{asymptotic}

We can gain insight into the complicated expressions for the dephasing and unitary-phase factors above by looking at their asymptotic forms. The full expressions for a single mode $k$ are shown in Fig.~\ref{fig:Factors} in the underdamped $\Gamma < \omega$ and overdamped regimes $\Gamma \gg \omega$.
Below, we discuss the asymptotic forms in these two regimes.

In the long-time limit, $\Gamma_k t \gg 1$ for all $k$ (i.e. for times $t$ much longer than any characteristic decay time $\Gamma_k^{-1}$), the oscillating transients die off, and the integrals become 
    \begin{align}
            {\cal{I}}_{\mathrm{Re}}(t; \omega_k, \Gamma_k) 
        &\rightarrow 2 |\eta_k|^2 \coth \big( \tfrac{ \beta  \omega_k }{ 2} \big) \frac{\Gamma_k }{\Gamma_k^2+4\omega_k^2}t
          \, , \label{eqIReLongTimeLimit}\\
        {\cal{I}}_{\mathrm{Im}}(t; \omega_k, \Gamma_k) 
        & \rightarrow -\frac{4 |\eta_k|^2 \omega_k t }{\Gamma_k^2+4\omega_k^2}
        \label{eqIImLongTimeLimit}
        \, .
\end{align}

\begin{figure}[t]\includegraphics[width=0.9\columnwidth]{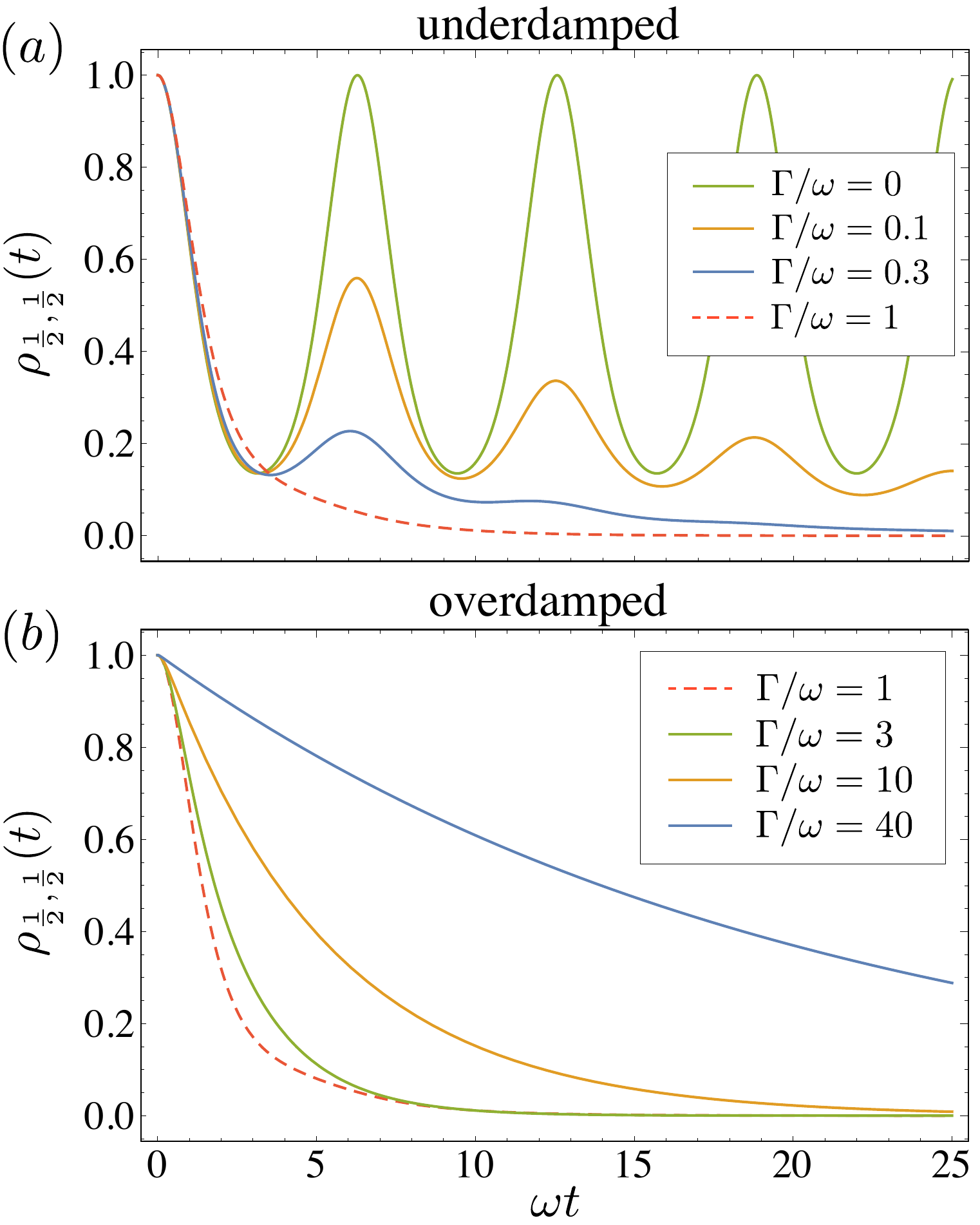}

\caption{
Decay of the off-diagonal coherences of a $j=\frac{1}{2}$ spin coupled to a single bosonic mode for varying decay rates. Plotted is the matrix element $\rho_{\frac{1}{2},\frac{1}{2}}(t)$
using Eq.~\eqref{eqRhoMMprimeSingleSpin} with $\rho_{\frac{1}{2},\frac{1}{2}}(0) = 1$.
Parameters are the same as in Fig.~\ref{fig:Factors}.
(a) Underdamped ($\Gamma < \omega$) and (b) overdamped ($\Gamma > \omega$ regimes).
The transition from underdamped, highly non-Markovian dynamics to simpler, overdamped dynamics is evident in the disappearance of oscillations.
In the underdamped regime, the decay rate of the spin coherence {\emph{increases}} with increasing decay rate $\Gamma$, while in the overdamped regime the decay rate of the spin coherence {\emph{decreases}} with increasing decay rate $\Gamma$. This can be seen from Eq.~(\ref{eqIReLongTimeLimit}), which shows that in the underdamped case the decay rate of the spin coherences scales proportional to $\Gamma$, while in the overdamped case the decay rate scales proportional to $\Gamma^{-1}$. This is a manifestation of the quantum Zeno effect; when the information the environment 
gains about the spin is lost fast enough, the spin state is frozen and does not decohere.
}
\label{fig:SingleSpinUnderAndOverDamped}
\end{figure}

There are two parameter regimes of interest for the spin dynamics. 
The first is the underdamped case, where $\Gamma_k < \omega_k$ for all $k$. The asymptotic expressions above become
    \begin{align}
        {\cal{I}}_{\mathrm{Re}}(t; \omega_k, \Gamma_k) 
        &\rightarrow |\eta_k|^2 \coth \big( \tfrac{ \beta  \omega_k }{ 2} \big) \frac{\Gamma_k}{2\omega_k^2}t 
          \, , \\
        {\cal{I}}_{\mathrm{Im}}(t; \omega_k, \Gamma_k) 
        & \rightarrow - |\eta_k|^2  \frac{1}{\omega_k} t  .
\end{align}
The magnitudes of spin coherences oscillate at frequency $|
\eta_k|^2/\omega_k$ while experiencing damping at a rate proportional to $\Gamma_k$.
This behavior is shown in the upper panel of Fig.~\ref{fig:SingleSpinUnderAndOverDamped} for the case of a spin-$\frac{1}{2}$ coupled to a single mode. Note that these oscillations arise from the non-Markovian dephasing-type dynamics generated by ${\cal{I}}_{\mathrm{Re}}$. In fact, for spin-$\frac{1}{2}$, the unitary-type oscillations generated by ${\cal{I}}_{\mathrm{Im}}$ vanish.

The second regime of interest is the overdamped case, where $\Gamma_k \gg \omega_k$ for all $k$, in which the magnitude of the spin coherences monotonically decreases. 
This behavior is due to the fact that the integral factors become
\begin{subequations} \label{eq:overdampedrates}
\begin{align}
    {\cal{I}}_{\mathrm{Re}}(t; \omega_k, \Gamma_k) 
        &\rightarrow 2 |\eta_k|^2 \coth \big( \tfrac{ \beta  \omega_k }{ 2} \big) \frac{1}{\Gamma_k} t
        \label{eq:overdampedRerates}
 \\
    {\cal{I}}_{\mathrm{Im}}(t; \omega_k, \Gamma_k) &\rightarrow 4 |\eta_k|^2 \frac{\omega_k}{\Gamma_k^2} t \approx 0 .
    \label{eq:overdampedImrates}
\end{align}
\end{subequations}
The spin coherence experiences no coherent oscillations and dephases at a rate that is \emph{inversely} proportional to the vibrational decay rate. 
Thus, in the overdamped regime, larger vibrational decay rates serve to preserve spin coherences. 
This can be interpreted in terms of a quantum Zeno effect: the modes measure the spin and then immediately discard the information into the environment --- similar to rapid projective spin measurements. 
To minimize decoherence of the spin, one desires weak spin-mode couplings $\eta_k$ and fast decay from the bosonic modes to their their baths.
Overdamped behavior is shown in the lower panel of Fig.~\ref{fig:SingleSpinUnderAndOverDamped}.




As a final note, we point out that the map in Eq.~\eqref{eqRhoMMprimeSingleSpin} (as well as the others like it throughout this work) can be used to describe the reduced Schr\"{o}dinger-type dynamics of any reduced-spin operator, not just density matrices, by expressing the operator in the $\hat{j}_z$ eigenbasis. Heisenberg-picture dynamics can be found simply by applying the propagator accordingly and following the same procedure.

\section{Dissipative protection of the symmetric subspace for two spins}
\label{sec:DissipativeProtectionOfCoherences}

We now explore how vibrational dissipation can have a protective effect on the coherences between many spins. We take each to be spin $j = \frac{1}{2}$ with eigenstates $\ket{m=\pm \frac{1}{2}}$, where the label giving spin $j=\frac{1}{2}$ is suppressed for brevity. Thus, the operator $\hat{j}^{(n)}_z$ for the $n$th spin satisfies 
    \begin{align}
        \hat{j}^{(n)}_z \ket{\pm \tfrac{1}{2}} = \pm \tfrac{1}{2} \ket{\pm \tfrac{1}{2}}
        \, .
    \end{align}
We consider a collection of such spins, each of which has local dynamics described by the spin-boson Hamiltonian in Eq.~\eqref{eqSpinMutipleHOModeCoupling1} (for $j = \frac{1}{2}$).  This physical setting
is motivated by two-level emitter defects in solids, where an electronic excitation deforms the surrounding crystal lattice, thus coupling to localized vibrational modes~\cite{BetzTorrBien14}. More details on this connection are given in Sec.~\ref{sec:NVExample}.


We focus on the pedagogical case of two spin-$\frac{1}{2}$ particles. Each spin couples to a quadrature of \emph{single} local vibrational mode. This means that each spin-mode pair is described by the spin-boson Hamiltonian in Eq.~\eqref{eqSpinMutipleHOModeCoupling1}, with just one term in the sum over $k$. Including a coupling between the two local vibrational modes of strength $\kappa$, the Hamiltonian describing this situation is given by
    \begin{align}
       \hat{H} 
       &=  \omega_0 \big( \hat{v}_1^\dagger \hat{v}_1 + \hat{v}_2^\dagger \hat{v}_2 + 1\big) + \kappa \big(\hat{v}_1^\dagger \hat{v}_2 + \hat{v}_1 \hat{v}_2^{\dagger}   \big) \nonumber \\
        &+ \eta \big[ \hat{j}_z^{(1)} \otimes (\hat{v}_1 + \hat{v}^{\dagger}_1) + \hat{j}_z^{(2)} \otimes (\hat{v}_2 + \hat{v}^{\dagger}_2) \big]  .
        \label{eqHTwoSpinsOneVibronic}
    \end{align}

The distributed (nonlocal) vibrational eigenmodes are given by symmetric and antisymmetric combinations of the local ones, 
    \begin{equation}
        \hat{v}_\pm = \tfrac{1}{\sqrt{2}} \left( \hat{v}_1 \pm \hat{v}_2 \right)    ,
    \label{eqSymmetricAndAntisymmetricModeDefinition}
    \end{equation}
whose eigenfrequencies are split by the coupling $\kappa$,
    \begin{equation}
        \omega_\pm \coloneqq \omega_0 \pm 2 \kappa  .
    \end{equation}
In this basis, the Hamiltonian can be rewritten as
    \begin{align}
        \hat{H} =  \omega_+ \hat{v}^{\dagger}_+ \hat{v}_+ +  \omega_- \hat{v}^{\dagger}_- \hat{v}_-
        + \eta \hat{J}_z \otimes \hat{x}_+ + \eta \hat{A}_z \otimes \hat{x}_- . 
        \label{normalmodeHam2spins}
    \end{align}
where $\hat{x}_\pm \coloneqq \frac{1}{\sqrt{2}} (\hat{v}_\pm + \hat{v}_\pm^\dagger)$ are the distributed-mode position quadrature operators. Coupled to the symmetric and anti-symmetric distributed modes are the (Hermitian) symmetric and antisymmetric collective operators,
    \begin{align}
        \hat{J}_z &\coloneqq \hat{j}_z^{(1)} + \hat{j}_z^{(2)} \\
        \hat{A}_z &\coloneqq \hat{j}_z^{(1)} - \hat{j}_z^{(2)}  .
    \end{align}
The eigenstates of collective spin operator $\hat{J}_z$ are the coupled angular momentum states $\ket{J,M}$, satisfying
    \begin{subequations}
    \begin{align}
        \hat{J}_z \ket{J,M} &= M \ket{J,M}, \\
        \hat{J}^2 \ket{J,M} &= J(J+1) \ket{J,M} ,
    \end{align}
    \end{subequations}
where $\hat{\mathbf{J}}$ is the total spin operator ($\hat{J}^2 = \hat{\mathbf{J}}\cdot\hat{\mathbf{J}} $). 
Relations between the local-spin and collective spin bases are given in Appendix~\ref{AppendixProjs}.

The actions of these two collective operators in the coupled-spin basis, where two spin half systems are treated as a collective spin-1 and a spin-0 particle, are
    \begin{align}
        \hat{J}_z &= \sum_{J=0}^1 \sum_{M=-J}^J M \ketbra{J,M}{J,M} \\
        \hat{A}_z &= 2 \big( \ketbra{1,0}{0,0} + \ketbra{0,0}{1,0} \big)  ,
    \end{align}
This description makes it clear that the $\hat{A}_z$ operator couples the $J=1$ and $J=0$ subspaces without adding or removing spin excitations (indicated by no change in the $M$ label). Meanwhile, collective spin operators such as $\hat{J}_z$ are block-diagonal in the coupled-spin basis, which separates them into their irreducible representations~\cite{Sakurai2010}; here $\hat{J}_z = \hat{J}^{(1)}_z \oplus \hat{J}^{(0)}_z$, where $\hat{J}^{(i)}_z$ is the spin-$i$ irreducible representation.

Each of the symmetric and antisymmetric vibrational modes decays into its own thermal bath, giving a master equation,
\begin{equation}
\frac{d}{dt} \hat{\rho} =  -i \big[ \hat{H}, \hat{\rho} \big] +  \mathcal{D}^{\text{th}}_+[\hat{\rho}]  +  \mathcal{D}^{\text{th}}_-[\hat{\rho}] 
\end{equation}
where the thermal dissipators are defined in Eq.~(\ref{dissipator}) and have respective decay rates $\Gamma_\pm$. Note that this is different from the above case, where each vibrational mode decays locally.
Although the spins are not directly coupled to one another, their local vibrational modes may be (for $\kappa \neq 0$), and further, the vibrational modes decay in a collective symmetric and antisymmetric fashion. These give rise to effective spin-spin coupling, which can be seen in the evolution of the reduced spin state $\hat{\rho}_\text{spin}(t) = \mathrm{Tr}_V [\hat{\rho}(t)]$.
The matrix elements in the local-spin basis,
$\rho_{m_1, m_2}^{m_1', m_2'}(t) \coloneqq \bra{m_1, m_2} \hat{\rho}_\text{spin}(t) \ket{m'_1, m'_2} $, evolve according to
    \begin{align}
        \rho_{m_1, m_2}^{m_1', m_2'} (t) &= \rho_{m_1, m_2}^{m_1', m_2'}(0)  \exp \Big\{  \nonumber \\
        & i [(m_1^2 + m_2^2)- (m_1'^2 + m_2'^2)] \mathcal{I}_{\mathrm{Im}}(t; \omega_+, \Gamma_+) \nonumber \\
        & + i [(m_1^2 - m_2^2) - (m_1'^2 - m_2'^2)] {\cal{I}}_{\mathrm{Im}}(t; \omega_-, \Gamma_-)  \nonumber \\
        &  - [(m_1 + m_2) - (m_1' + m_2')]^2 {\cal{I}}_{\mathrm{Re}}(t; \omega_+,  \Gamma_+) \nonumber \\
        & - [(m_1 - m_2) - (m_1' - m_2')]^2 {\cal{I}}_{\mathrm{Re}}(t; \omega_-,  \Gamma_-) \Big\}  .
    \label{eqRhoM1M2SymAndAntiSym}
    \end{align}
The terms associated with the symmetric vibrational mode have sums of $m_1$ and $m_2$ and those associated with the antisymmetric vibrational mode have differences of $m_1$ and $m_2$.


\subsection{Preserving the symmetric subspace}
We are interested in preserving the symmetric spin subspace where the symmetric Dicke states lie. In the local spin basis, the diagonal elements of the density matrix that describe local-spin populations do not evolve. However, the collective spin populations do, since the states $\ket{J,M=0}$ contain local-spin coherences. The symmetric subspace is described by the rank-3 projector
    \begin{align}
        \hat{P}_\text{sym} \coloneqq \sum_{M=-1}^1 \hat{P}_{1,M} ,
    \end{align}
where $\hat{P}_{J,M} \coloneqq \ketbra{J,M}{J,M}.$
Because dephasing is diagonal in the local spin basis, the projectors $\hat{P}_{1,1}$ and $\hat{P}_{1,-1}$ are stationary in time. Population only leaves the symmetric subspace through the state $\ket{J=1,M=0}$. 
Details can be found in Appendix~\ref{AppendixProjs}. 

\begin{figure}[tb!]
\includegraphics[width=0.9\columnwidth]{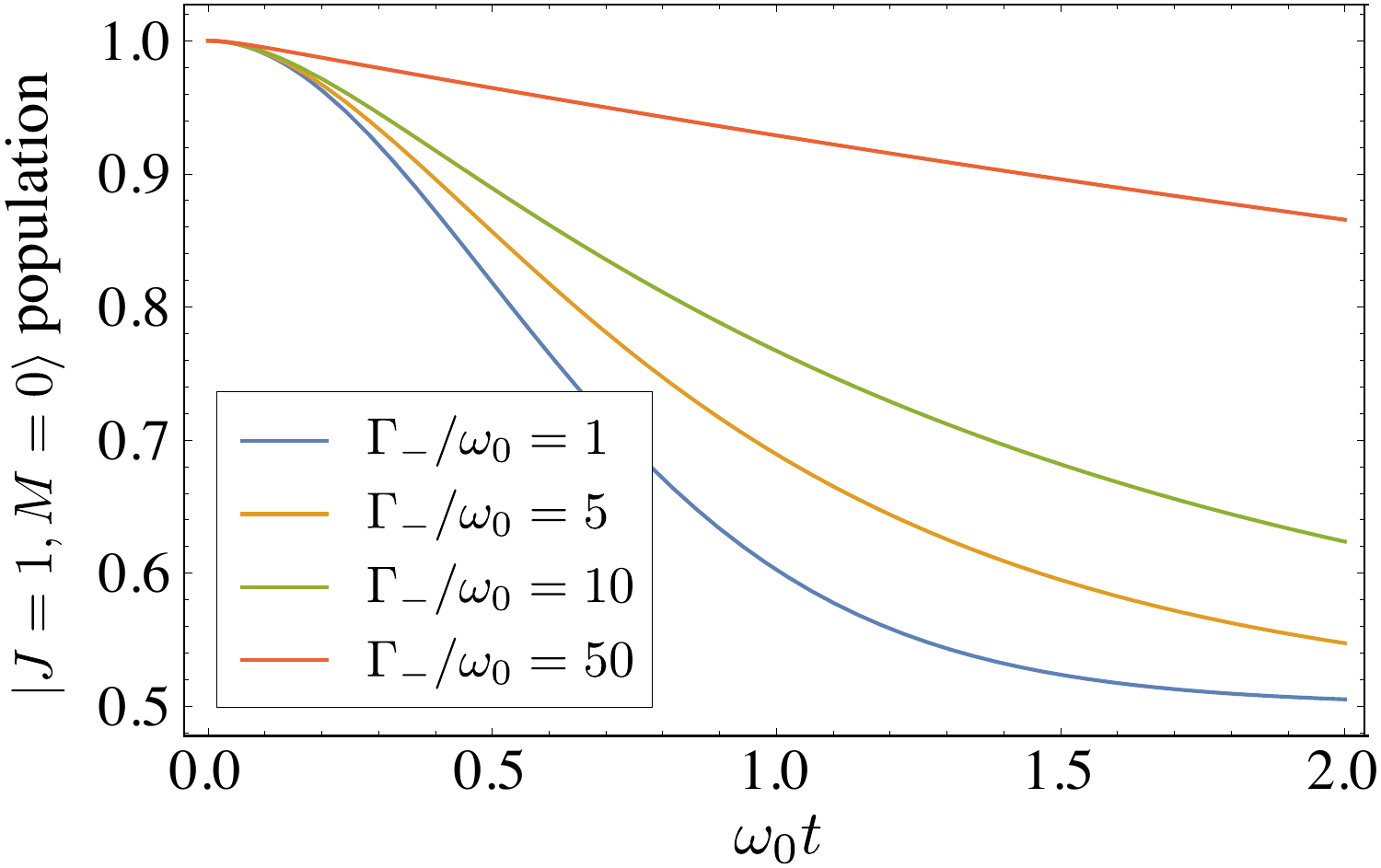}
\caption{Decay of the symmetric two-spin state $\ket{J=1,M=0}$ 
as a function of time for various $\Gamma_-$.
The parameters are: $\eta/\omega_0 = 1$, $\kappa =0$, and
$\beta \rightarrow \infty$. The curves show population decaying out of the initial state.
Lost population is incoherently pumped into the anti-symmetric state $\ket{J=0,M=0}$. 
The quantum Zeno effect is apparent, as the population is preserved for longer times  by increasing $\Gamma_-$.
}
\label{fig:CrossModeCouplingPreservesSymmetricSubspace2}
\end{figure}

Using Eq.~\eqref{eqRhoM1M2SymAndAntiSym}, the projector onto this state evolves as
 \begin{align}
        \hat{P}_{1,0}(t) & = 
        \frac{1}{2} \big[ 1 + e^{-4 \mathcal{I}_\text{Re}(t;\omega_-, \Gamma_-) } \big] \hat{P}_{1,0} \nonumber \\
        &+ \frac{1}{2} \big[ 1 - e^{-4 \mathcal{I}_\text{Re}(t;\omega_-, \Gamma_-) } \big] \hat{P}_{0,0} \label{symmetricdecay} ,
    \end{align}
revealing that population lost from $|J=1,M=0\rangle$ moves to $|J=0,M=0\rangle$. 
The result is that the overlap of the symmetric-subspace projector decays over time,
    \begin{align}
        \text{Tr} \big[\hat{P}_\text{sym}(t) \hat{P}_\text{sym}(0) \big] 
        & = 2 +  \frac{1}{2} \left( 1 + e^{-4 \mathcal{I}_\text{Re}(t;\omega_-, \Gamma_-) }\right) .
        \label{eqSymmetricSubspaceOverlapProjector}
    \end{align}
Crucially, the decay exponent scales as $\Gamma_-^{-1}$ in the overdamped limit where $\Gamma_- \gg \omega_-$ as shown in Eq.~(\ref{eq:overdampedRerates}). This means that as the antisymmetric vibrational-mode decay rate increases, the rate at which population leaves the symmetric subspace decreases. This is a manifestation of the quantum Zeno effect, wherein a quantum system with support on a subspace $A$ and that is coherently coupled to a second subspace $B$ that is measured at a fast rate, is effectively decoupled from $B$ thus preserving coherence in $A$. 
Symmetric state preservation with increasing $\Gamma^-$ is illustrated in  Figure~\ref{fig:CrossModeCouplingPreservesSymmetricSubspace2}.

Note that Eq.~(\ref{eqSymmetricSubspaceOverlapProjector}) does not involve the symmetric decay rate $\Gamma_{+}$ at all. Although we do not explore the effect here, symmetric-mode decay can act as dephasing within the symmetric subspace. This reduces coherences between the off-diagonal elements in the collective basis, but unlike the antisymmetric decay $\Gamma_-$, it does not move population out of that symmetric subspace.

\begin{figure}[tb!]
\includegraphics[width=0.9\columnwidth]{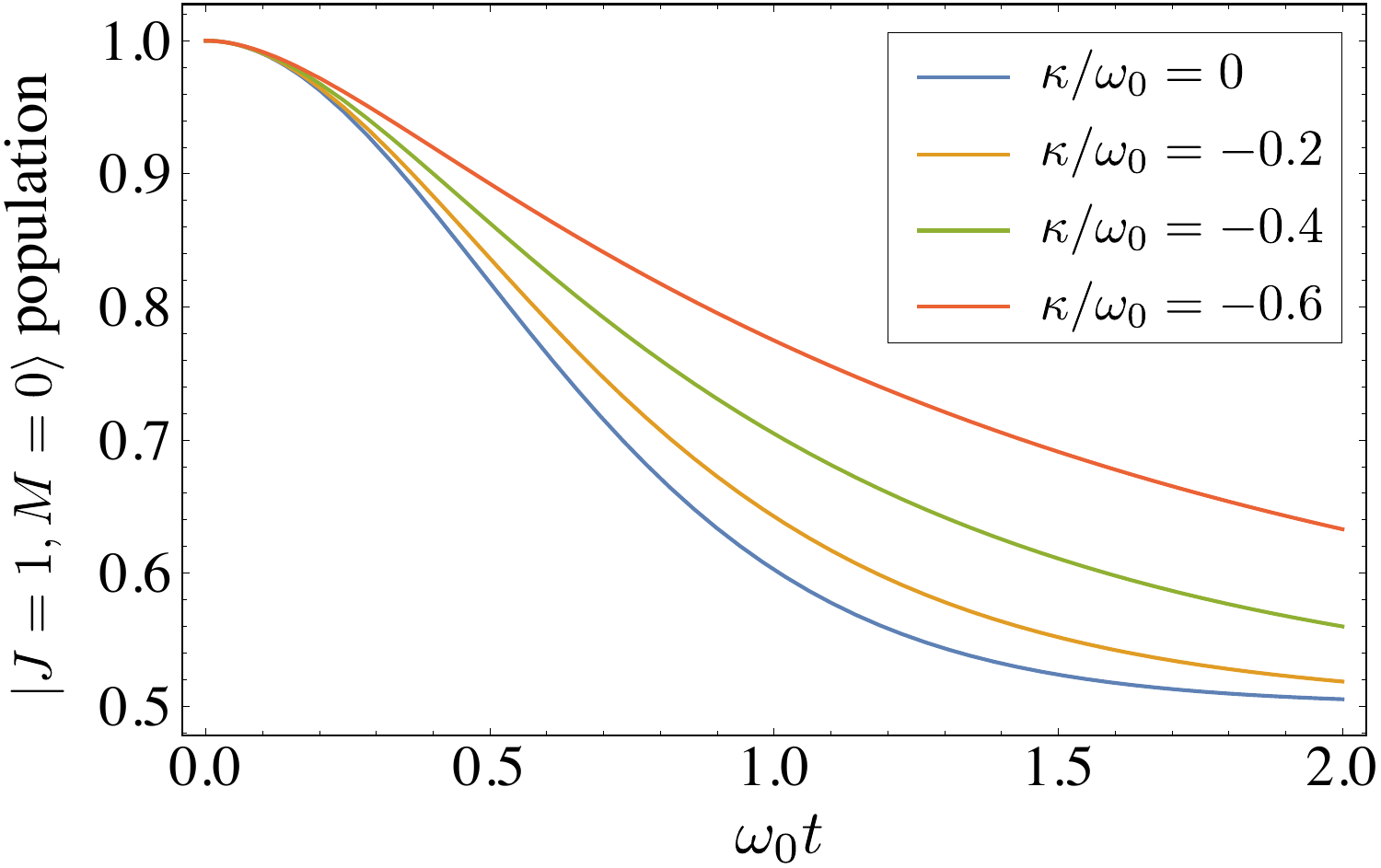}
\caption{
Influence of inter-mode coupling $\kappa$ on symmetric-state decay.
Decay of the symmetric two-spin state $\ket{J=1,M=0}$ induced by coupling to the decaying bosonic modes.
The parameters are: $\eta/\omega_0 = 1 $, $\Gamma/\omega_0 = 1$, 
$\Gamma_{\pm} /\omega_0=  (\omega_{\pm}/\omega_0)^3$, 
and $\beta \rightarrow \infty$. 
}
\label{fig:CrossModeCouplingPreservesSymmetricSubspace1}
\end{figure}

The role of the coupling $\kappa$ between the bosonic modes is to shift the energies of the eigenmodes and hence their decay rates. Since we assume the bosonic mode decay arises due to Lindblad type dynamics with a Markovian bath, we can write the decay rates according to Fermi's golden rule $\Gamma(\omega)=2\pi D(\omega)|g|^2$ where $g$ is some fundamental coupling rate between initial and final bosonic mode states and $D(\omega)$ is the density of states at the energy of the final states. For the Debye model of coupling between phonons in three dimensional systems, the density of states scales like $D(\omega)\propto \omega^3$ as does $\Gamma(\omega)$. Hence for this simple model of two vibrational modes, we have 
\begin{equation}
\frac{\Gamma_{\pm}}{\Gamma}=\Big(\frac{\omega\pm 2\kappa}{\omega_0}\Big)^3 ,
\label{decayratesmod}
\end{equation}
where $\Gamma$ is the mode decay rate in absence of interactions.
Lifting the mode degeneracy via nonzero inter-mode coupling will decrease ($\kappa > 0$) or increase ($\kappa < 0$) the antisymmetric mode decay rate relative to the symmetric mode decay rate as shown in (\ref{decayratesmod}). The latter case is useful for preserving the symmetric subspace as illustrated in Figure~\ref{fig:CrossModeCouplingPreservesSymmetricSubspace1}. This protection is accompanied by lower symmetric decay rates, which reduces the spin dephasing within the collective spin subspace.


\section{Using inter-mode couplings to alter the environmental structure}
\label{sec:structured}

\begin{figure}[t]
\includegraphics[width=0.8\columnwidth]{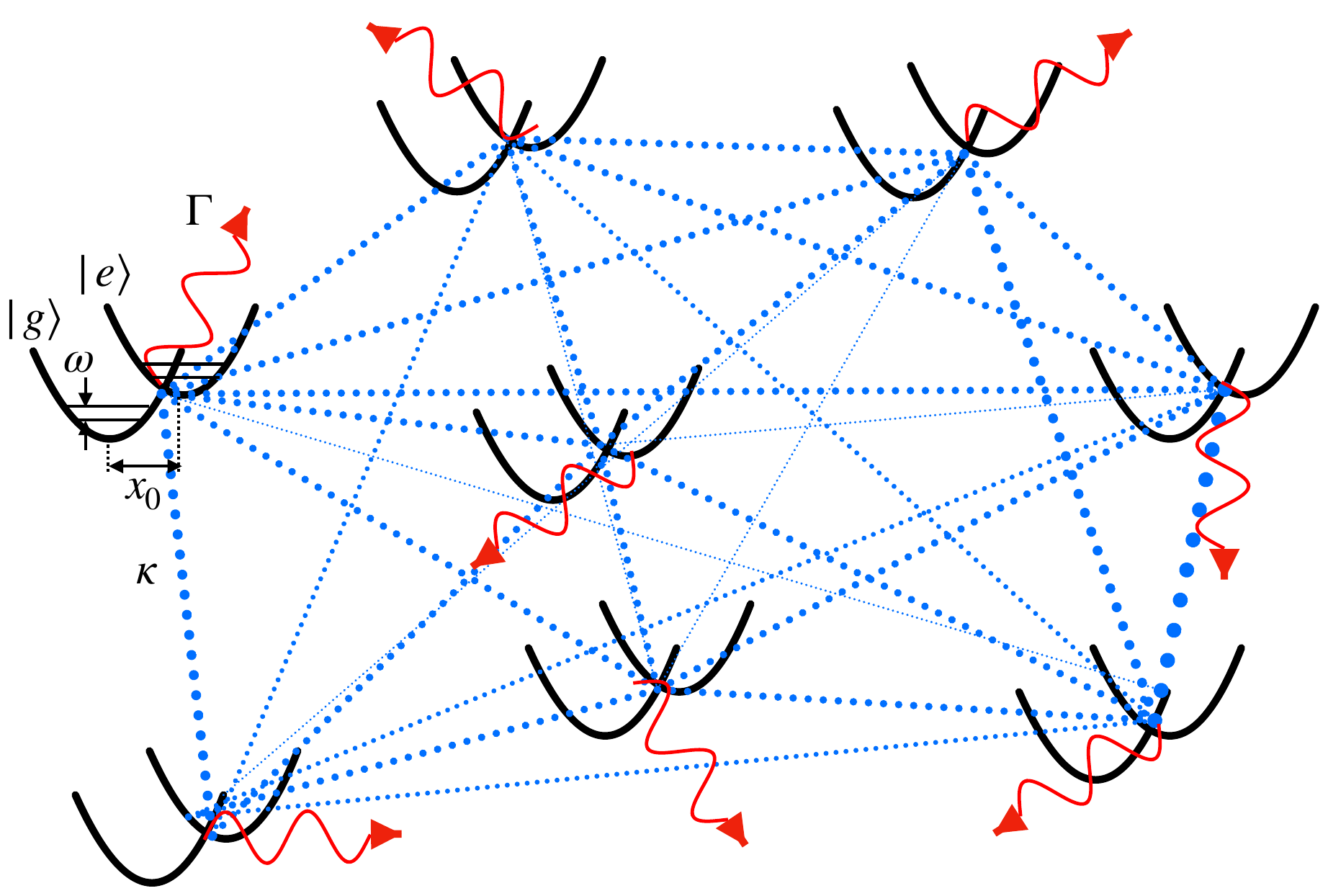}
\caption{\label{fig:Setup} 
Depiction of multiple spins whose local modes are coupled together, illustrated in the context of two-level defects in solid state (see Sec.~\ref{sec:NVExample}). Each pseudo-spin couples in a state-dependent manner to local bosonic modes, \emph{e.g.} a local vibronic mode, shown as a state dependent displacement of the harmonic oscillator potential by an amount $x_0=2x_{\rm rms}\eta/\omega$, where $x_{\rm rms}$ is ground state width and $\omega$ the energy of the mode. Further, the modes are coupled to each other with potentially different strengths $\kappa_{ij}$ (blue dashed lines), and also experience decay into a bath of extended thermal modes, e.g. phonons, at rate $\Gamma$ (red wavy lines).}
\end{figure}

In this section, we consider an extension of the spin-dephasing Hamiltonian analysed in the earlier sections, where the now the bosonic modes are coupled to each other and there are many spins. We will show that when the modes have a specific structure, a single normal mode arises that couples to the symmetric spin subspace. Then, if the symmetric subspace can be preserved for some time, that subspace may be treated like a large spin of size $J = N/2$ and the results of the previous sections can be applied.
Specifically, we consider the effects of a structured environment where the modes couple together as in Fig.~\ref{fig:Setup}. We will show that for certain types of inter-mode coupling, discussed below, a large spectral gap opens between the (near)-symmetric mode and all other normal modes. 
The energy splitting can be used to address this mode. When the modes are coupled to spins with a spin-boson coupling, selectively addressing the symmetric mode can be useful for quantum information protocols. For example, when the mode decay rates are small or absent, one can use this gap to engineer a geometric phase gate~\cite{sorenson1999} that gives rise to nonlinear interactions between the spins.

Consider the $N$-spin generalization of the spin-boson Hamiltonian in Eq.~\eqref{eqHTwoSpinsOneVibronic} with each spin coupled to a single mode,
\begin{align}
       \hat{H} 
       &= \hat{H}_m 
        +\sum_{k=1}^N  \hat{j}_z^{(k)} \otimes (\eta_k \hat{v}_k + \eta^*_k \hat{v}^{\dagger}_k )  .
        \label{eqHNSpinsOneVibronic}
\end{align}
where the mode-wise part of the Hamiltonian is
\begin{align} 
\label{modeHam}
    \hat{H}_m = \omega_0  \sum_{k=1}^N ( \hat{v}_k^\dagger \hat{v}_k + \tfrac{1}{2} ) + \sum_{k \neq k'} \kappa_{k,k'} \hat{v}_k^\dagger \hat{v}_{k'},
\end{align}
The mode-wise couplings $\kappa_{k,k'}$ can be grouped into a matrix $\boldsymbol{\kappa}$. Diagonalizing $\hat{H}_m$ (by diagonalizing $\boldsymbol{\kappa}$) gives rise to $N$ normal bosonic modes $\hat{w}_{k'}$ with associated eigenfrequencies $\lambda_{k'}$.\footnote{Even though $k$ and $k'$ are dummy indices, we denote them differently for clarity.} This allows the mode-wise part of the Hamiltonian to be written as
\begin{equation}
    \hat{H}_m =\sum_{k'=1}^N \lambda_{k'} (\hat{w}^{\dagger}_{k'}\hat{w}_{k'} + \tfrac{1}{2} ) .
\end{equation}
Formally, the eigenmodes are constructed from the original modes by
\begin{equation}
    \hat{w}_{k'} = \sum_{k=1}^N \langle{\lambda_{k'}}\ket{k} \hat{v}_k .
\end{equation}
where $\ket{\lambda_{k'}}$ is an eigenvector of the matrix $\boldsymbol{\kappa}$ (not a Hilbert-space vector), and $\ket{k}$ is a unit vector with all zero entries except at position $k$. 
Using the normal modes, the full spin-boson Hamiltonian in Eq.~\eqref{eqHNSpinsOneVibronic} can be written
\begin{equation}
\hat{H}=\sum_{k=1}^N \lambda_{k} \big( \hat{w}^{\dagger}_{k}\hat{w}_{k}+\tfrac{1}{2} \big)
+
\sum_{k} ( \hat{O}_{k} \otimes \hat{w}^{\dagger}_{k} + \hat{O}_{k}^\dagger \otimes \hat{w}_{k} ) .
\label{fullHam}
\end{equation}

Each normal mode couples to collective operators in the spin degree of freedom,
    \begin{align}
        \hat{O}_k \coloneqq \sum_j \langle{j}\ket{\lambda_k} \eta_j \, \hat{j}_z^{(j)}.
    \end{align}
In general, these are not proportional to collective spin operators. However, when $\eta_j = \eta$, i.e. the coupling of the spins to local modes is homogeneous, and one of the normal modes is a symmetric mode defined by 
\begin{equation}  \label{symmetricmode}
\hat{w}_s \coloneqq \frac{1}{\sqrt{N}}\sum_{k=1}^N\hat{v}_k.
\end{equation}
Then one of the spin system operators is $\hat{O}_s = \frac{\eta}{\sqrt{N}} \hat{J}_z$, where
\begin{equation} \label{eq:Jz}
        \hat{J}_z \coloneqq \sum_{j=1}^N \hat{j}_z^{(j)},
\end{equation}
is a collective spin operator.

We now discuss situations where the symmetric mode
with eigenfrequency $\lambda_s$
couples to $\hat{J}_z$ and how this can be exploited for nonlinear interactions in the collective spin.

\subsection*{Structured modewise coupling to isolate the symmetric normal mode}
\label{sec:StucturedModewiseCoupling}

The normal mode spectrum is determined by $\boldsymbol{\kappa}$. We assume the spin-mode coupling rates are uniform, $\eta_j = 
\eta$, and consider $\boldsymbol{\kappa}$ with certain structure, described below, where the symmetric mode $\lambda_s$ is gapped from the rest of the eigenmodes.
 In this scenario the Hamiltonian Eq.~(\ref{fullHam}) can be written
 \begin{equation}
\hat{H}=\sum_{k\neq s} \lambda_k \big( \hat{w}^{\dagger}_k\hat{w}_k+\tfrac{1}{2} \big)
+
\sum_{k\neq s} ( \hat{O}_k \otimes \hat{w}^{\dagger}_k + \hat{O}_k^\dagger \otimes \hat{w}_k )+\hat{H}_s
\end{equation}
where the dynamics of the symmetric mode is 
\begin{equation}
\hat{H}_s= \lambda_s \big( \hat{w}^{\dagger}_s\hat{w}_s + \tfrac{1}{2} \big) +
\frac{\eta}{\sqrt{N}} \hat{J}_z \otimes (\hat{w}^{\dagger}_s + \hat{w}_s)  .
\label{symHam}
\end{equation}

Although we are most interested in the symmetric spin space, note that $\hat{J}_z$ operator has support over the entire Hilbert space of the spins (dimension $2^N$). Thus, it is not simply an operator for the single $2J + 1$-dimensional symmetric space (total spin $J = N/2$) but acts all the irreducible representations of angular momentum and their multiplicities~\cite{BaragiolaSpinNoiseDephasing2010}.
However, $\hat{J}_z$ does not mix these subspaces, so $\hat{H}_s$ likewise does not mix spin-subspaces of different permutation symmetry. This is important because the collective spin state is often prepared in the symmetric subspace.

Now using the spectral resolvability of the symmetric mode it is possible to perform quantum control on the dynamics to effectively restrict evolution to the symmetric subspace. 
Consider a dynamical decoupling pulse sequence where periodically with period $T$ the unitary $\hat V=e^{i\pi \hat{J}_x}e^{i\pi \hat{w}_s^{\dagger}\hat{w}_s}$ is applied, i.e. a composition of an on-site bit flip on the qubits and $\pi$ phase shift on the symmetric mode. This describes a bang-bang decoupling sequence and if it is done fast relative to the coupling strength, i.e. 
$T^{-1}\gg \eta/\sqrt{N}$, then the effective evolution will be restricted to that generated by $\hat{H}_s$ \cite{PhysRevA.58.2733}. Such evolution can be used to generate spin squeezing. For example, in the limit that the mode decay rates go to zero $\Gamma_k\rightarrow 0$ and when the overall time of evolution for the dynamically decoupled sequence satisfies $\omega_s t= r\pi$ for $r \in \mathbb{N}$, then by Eq.~(\ref{nodecayfactors}) the evolution acts on the spins alone according to the unitary $\hat U =e^{-i \chi \hat{J}_z^2}$ with 
$\chi=\frac{r\pi\eta^2}{N\omega_s^2}$.
This is similar to a control scheme considered by Chaudry and Gong~\cite{Chaudhry2013}, who proposed using quantum control on the symmetric subspace of many spins to mitigate spin-mode correlations that can be problematic when the bath correlation function is not known.

\begin{figure}[tb!]
\includegraphics[width=0.9\columnwidth]{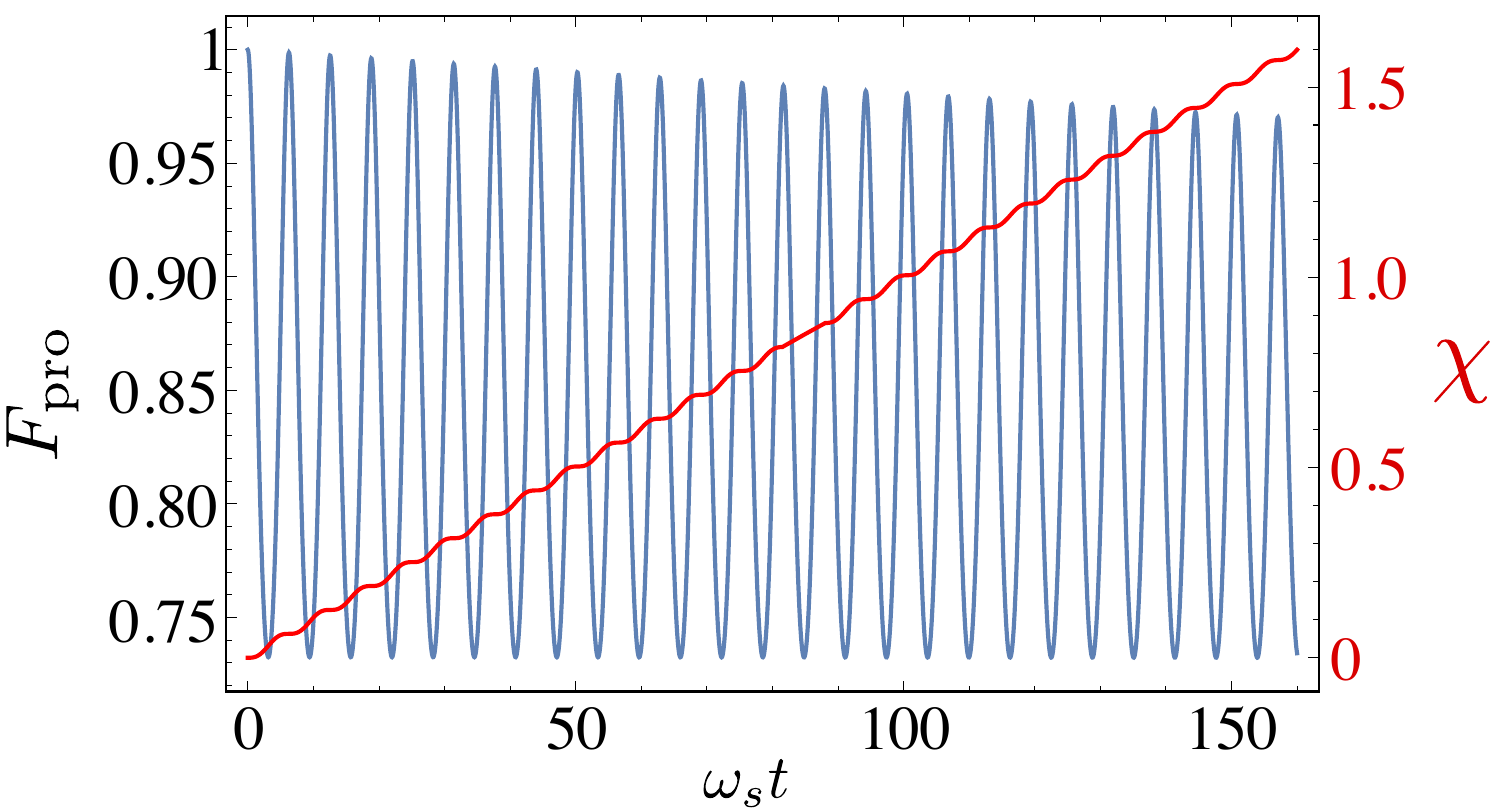}
\caption{
Process fidelity (blue) $F_{\rm pro}(\mathcal{E},U)$ for a target spin squeezing unitary $\hat U=e^{-i\chi \hat{J}_z^2}$ due to a collective spin with angular momentum $j=5$ evolving according to the damped, pure-dephasing model with a single mode $\omega_s$. The system parameters are in the underdamped regime, $\eta/\omega_s=0.1$, $\Gamma/\omega_s=10^{-3}$, and $\beta=10$. The accumulated coherent phase $\chi$ is plotted in red. At $t=157.08/\omega_s$, $\chi=\pi/2$ with $F_{\rm pro}=0.9703$.
}
\label{fig:ProcessFid}
\end{figure}

We can quantify how well one can perform the spin squeezing unitary via the process fidelity. Consider a target process $\mathcal{U}(\rho)= \hat U \hat \rho  \hat{U}^{\dagger}$ where $\hat U=e^{-i\hat{J}_z^2 \mathcal{I}_{\rm Im}(t;\omega_s,\Gamma)}$ originates in the unitary-phase part of the full process $\mathcal{E}$, given in terms of spin matrix element evolution in Eq.~(\ref{eqRhoMMprimeSingleSpin}), where we consider a collective spin state confined to the symmetric subspace. 
Due to the fact that the dephasing part of the dynamics commutes with the unitary-phase part, we can write 
$\mathcal{E}=\mathcal{E'}\circ \mathcal{U}$, where $\mathcal{E'}(\ket{J,M}\bra{J,M'})=e^{-(m-m')^2\mathcal{I}_{\rm Re}(t;\omega_s,
\Gamma)}\ket{J,M}\bra{J,M'}$.
The quality of this process is quantified by the process fidelity, which in the case of a target unitary is proportional to the average fidelity over pure states of the spin system \cite{PhysRevA.71.062310}. For this process it is
\begin{align}
    F_{\rm pro}(\mathcal{E},\mathcal{U})
    &=F_{\rm pro}(\mathcal{E'},\mathbb{I})\\
    &=\bra{\Phi^+}\hat{\rho}_{\mathcal{E'}}\ket{\Phi^+}.
\end{align}
where $\mathbb{I}$ is the identity channel, and $\mathcal{E'}$
is the dephasing map. In the second line, we express the process fidelity between these two in the Jamio\l kowski-Choi representation, where
according to channel-state duality~\cite{Fu2013}, each process corresponds to a state in a larger Hilbert space $\mathcal{H}_S\otimes \mathcal{H}_S'$ with $\mathcal{H}_{S'}$ being a copy of the Hilbert space of our spin $\mathcal{H}_S$ . 
The identity channel is given by the state
$\ket{\Phi^+} \coloneqq \frac{1}{\sqrt{2J+1}}\sum_{m=-J}^J \ket{J,M}_S \otimes \ket{J,M}_{S'}$, and the dephasing map is given by the state $\hat{\rho}_{E'}$.
Then, the process fidelity can be readily calculated as
\begin{equation}
    F_{\rm pro}(\mathcal{E},\mathcal{U})=\frac{1}{(2J+1)^2}\sum_{m,m'=-J}^J e^{-(M-M')^2 \mathcal{I}_{\rm Re}(t;\omega_s,
    \Gamma)}.
\end{equation}
Fig.~\ref{fig:ProcessFid} shows the process fidelity for a $J=5$ spin in the highly underdamped regime. We see in this case the oscillations of the process fidelity as the system periodically nearly decouples from the mode degree of freedom while the coherent phase increases essentially linearly with time. For the higher damping, the process fidelity degrades continuing into the extreme overdamped regime where the decay due to increased real part of the correlation function $\mathcal{I}_{\rm Re}$ goes to zero, but so too does the coherent phase.

We now demonstrate several conditions where such a gap between the symmetric mode and the other modes can arise.
We consider situations where $\boldsymbol{\kappa}$ is negative; i.e. all the mode-wise couplings satisfy $\kappa_{ij} <0$. 
Within this parameter regime, we study two types of structured mode-wise coupling. The first is uniform negative mode-wise coupling and the second is random (but negative) mode coupling. 



\subsubsection{Uniform coupling}

We begin with the case where the mode-wise couplings are all equal, $\kappa_{i,j} = \kappa$, making $\boldsymbol{\kappa}$ proportional to the unit matrix.
Diagonalizing $\boldsymbol{\kappa}$ gives $N-1$ degenerate modes of frequency $|\omega_0 - \kappa|$, and one privileged fully symmetric mode, Eq.~\eqref{symmetricmode}, with frequency $|\omega_0 + (N-1)\kappa|$. Note that when $\kappa < 0$, the symmetric mode is the lowest energy mode, and when $\kappa > 0$, it is the highest energy one.

\subsubsection{Random coupling}
We also assume the coupling is a perturbation to the bare mode coupling, specifically that the matrix $(\omega_0 {\bf 1}_N+\boldsymbol{\kappa}$) is positive.
The bare mode Hamiltonian can be written in terms of normal modes as 
\begin{equation}
\hat H=\sum_{j=1}^N \lambda_j \big( \hat{w}^{\dagger}_j\hat{w}_j+\tfrac{1}{2} \big)
\end{equation}
where $\{\lambda_i\}$ are the (increasing ordered) eigenvalues of the matrix, 
$\sqrt{(\omega {\bf 1}_N+\boldsymbol{\kappa})^2}$, and the eigenmodes $\hat{w}_j=\sum_{j,k}c_{j,k}\hat{v}_k$ are determined from the eigenvectors $\ket{\lambda_j}=\sum_k c_{j,k}\ket{k}$
of the matrix $\boldsymbol{\kappa}$. By the Frobenius-Perron theorem, there is a unique smallest eigenvalue $\lambda_1$ and the corresponding normal mode has strictly positive coefficients, $c_{1,j}> 0$. Furthermore, according to a theorem of F\"uredi and Komlos \cite{Furedi1981}, if the couplings are described by independent (not necessarily identically distributed) random variables, with a common bound and with common mean $\mathbb{E}[\kappa_{j,k}]=\mu<0$ and variance $\mathbb{E}[(\kappa_{j,k}-\mu)^2]=\sigma^2$, then the unique smallest eigenvalue satisfies
\begin{equation}
\lambda_1=\omega_0 -\frac{\sigma^2}{|\mu|} + \frac{1}{N} \sum_{j,k=1}^N\kappa_{j,k}+O(1/\sqrt{N})  ,
\end{equation}
and all the other eigenvalues are concentrated in the interval
$[\omega_0-c\sqrt{N},\omega_0+c\sqrt{N}]$,
where $c$ is any constant greater than $2\sigma$. This implies an expected spectral gap between the ground and second lowest energy modes of
\begin{equation}
    \mathbb{E}[\lambda_2-\lambda_1]=|\mu| N+\frac{\sigma^2}{|\mu|}-2\sigma\sqrt{N}+O(1/\sqrt{N}).
\end{equation}

Furthermore, the lowest energy mode is close to the fully symmetric mode. 
From Lemma 3 in Ref.~\cite{Furedi1981}, the ground-state eigenvector $\ket{\lambda_1}$ has overlap with the uniform state $\ket{v}=\frac{1}{\sqrt{N}}\sum_j \ket{j}$ of $\bra{\lambda_1}v\rangle>1-\frac{2\sigma^2}{N\mu^2}$ with probability $P >1-1/N$.


\begin{figure}[t!]
\includegraphics[width=0.9\columnwidth]{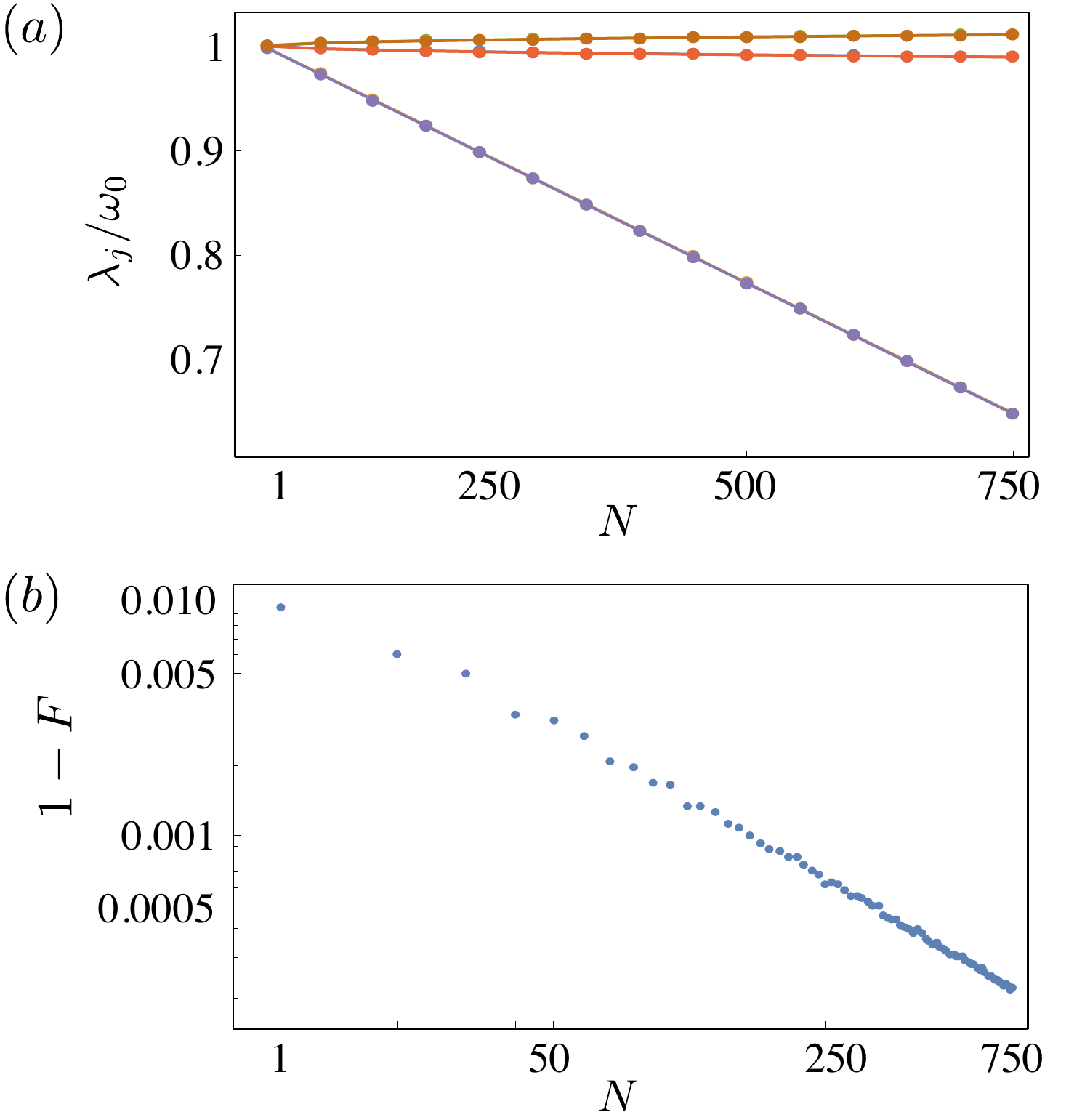}
\caption{Illustration of the separation of symmetric mode from the other modes for a collection of $N$ randomly coupled modes. (a) Shown is the lowest energy eigenmode, i.e. the symmetric mode $\hat{w}_s$ (blue), the second lowest (red) and the highest (orange) energy modes. Here the bare mode frequency is $\omega_0$ and the mode couplings are chosen randomly and uniformly in the interval $ [-\kappa,0]$, where $\kappa/\omega=10^{-3}$. For each eigenmode and value of $N$ two points are plotted (nearly overlapping on the plot), indicating the mean $\pm$ one standard deviation energy as calculated over $10$ realizations of random couplings. (b) Mean fidelity error of the lowest energy eigenmode to the fully symmetric mode, where $F=(\frac{1}{\sqrt{N}}\sum_{j=1}^N \bra{j}\lambda_1\rangle)^2$. }
\label{fig:DecayRates}
\end{figure}

Summarizing, for a collection of $N$ identical bosonic modes coupled to each other in an all-to-all manner, a gap develops in the eigenmodes with a low-energy symmetric mode. 
When the inter-mode couplings are random with but with common negative mean and a common variance, the ground energy mode is close to the fully symmetric mode with a fidelity error that falls off as $1/N$, and the energy gap is $O(N)$. This is illustrated in Fig.~\ref{fig:DecayRates}

\section{Connection to emitter defects in solid-state systems}
\label{sec:NVExample}

Our analysis has up to this point has been quite general in terms of spins coupled to bosonic modes. 
We now show the connection to simple models of solid-state systems where defects in crystals couple to vibrational lattice phonons.
Such a system describes, for example, nitrogen vacancy (NV$^{-}$) centers in diamond, a topic of intense interest in the quantum information processing and sensing communities~\cite{neumann2010,maletinsky2012,doherty_nitrogen-vacancy_2013,neuman2013, dohertyET2014, schirhagl2014,JuanBradBesg16, BradJohnBreu17,casola2018,bucher2019,michl2019,karim2020}.

We focus here on a simplified model of a defect in a crystal considered by Betzholz \emph{et al.}~\cite{BetzTorrBien14},
where the defect is a two-level electronic system with ground and excited states $\ket{g}$ and $\ket{e}$ with transition frequency $\Omega$. In the context of NV$^{-}$ centres, $\ket{g}$ and $\ket{e}$ correspond to the $^{3}A_2$ and $^{3}E$ electronic levels respectively. 
The solid-state lattice is locally deformed due to the electronic excited-state orbital, coupling each two-level electronic system to a vibrational mode with bare frequency $\omega$. 
The Hamiltonian governing the dynamics is given by~\cite{BetzTorrBien14}
	\begin{equation} \label{BertHam}
		 \hat{H}_\text{el-vib} = \Omega \hat{P}_e + \omega \hat{v}\dg \hat{v} + \eta \hat{P}_e \otimes ( \hat{v} + \hat{v}\dg )
		 \, ,
	\end{equation}
where $\hat{P}_e \coloneqq \op{e}{e}$ is the projector onto the electronic excited state, and $\eta$ is the vibronic coupling strength. 
In a crystal setting, the local vibrational mode $\hat{v}$ couples to longer range acoustic phonons, which can serve as a thermal bath with dissipator as in Eq.~\eqref{dissipator}. 

The reduced state of the electronic subsystem can be calculated using the solutions for the spin-boson model we derived in previous Sections. Additionally, our results regarding protection of coherences and symmetric subspaces, as well as the effect of randomized mode couplings, will also apply to these lattice defect systems. 

To see this, again we assume an initial joint state $\hat{\rho}(0) = \hat{\rho}^\text{el} (0) \otimes \hat{\rho}_V(0)$. The initial electronic state is arbitrary and has matrix elements $\rho_{jk} \coloneqq \bra{j} \hat{\rho}^\text{el} (0) \ket{k}$ for $j,k \in \{ g, e\}$, and $\hat{\rho}_V(0)$ is a (single-mode) thermal state, Eq.~\eqref{thermalstatek}.
As shown in Appendix \ref{ConnectHams}, the reduced spin state is given by (\ref{BertHam}) is
    \begin{eqnarray}
       \hat \rho^\text{el}(t) 
       &=& \rho_{gg} \op{g}{g} + \rho_{ge} e^{  -i {\cal{I}}_{\mathrm{Im}} - {\cal{I}}_{\mathrm{Re}} } \op{g}{e}  \nonumber \\
       &&+ \rho_{eg} e^{  i {\cal{I}}_{\mathrm{Im}} - {\cal{I}}_{\mathrm{Re}} }\op{e}{g}  + \rho_{ee} \op{e}{e} .
    \end{eqnarray}  
The dynamical phase and decay factors, $\cal{I}_\mathrm{Im}$ and $\cal{I}_\mathrm{Re}$, are shortened notation for those defined in Eqs.~\eqref{fancyintegrals}. Here, they are evaluated for the single frequency $\omega$ (since there is only a single mode). If the electronic defect couples to multiple modes, extension to this case is straightforward using by using their more general forms.

An excited defect or color centre can undergo optical decay on the timescale of a few nanoseconds \cite{doherty_nitrogen-vacancy_2013} as well as additional pure dephasing.
These two effects were not considered in the spin-boson model above, as we focused there on interactions between spins and bosonic modes (vibrational phonon modes in this context). To include these two additional processes, we add to our master equation Lindblad terms describing optical decay at a rate $\Gamma_{\text{op}}$ and additional dephasing at a rate $\Gamma_{\text{dp}}$.
As shown in Appendix~\ref{ConnectHams}, this results in a reduced electronic state given by 
    \begin{eqnarray}
       \hat \rho^\text{el}(t) 
       &=& (1 - \rho_{ee} e^{-\Gamma_\text{op} t}) \op{g}{g} \nonumber \\
       && + \, \rho_{ge} e^{-(\Gamma_{\text{dp}}+\frac{\Gamma_\text{op}}{2}) t} e^{  -i {\cal{I}}_{\mathrm{Im}} - {\cal{I}}_{\mathrm{Re}} } \op{g}{e} \nonumber \\
      &&  + \, \rho_{eg}e^{-(\Gamma_{\text{dp}}+\frac{\Gamma_\text{op}}{2}) t}  e^{  i {\cal{I}}_{\mathrm{Im}} - {\cal{I}}_{\mathrm{Re}} } \op{e}{g} \nonumber \\
       && \, + e^{-\Gamma_\text{op} t} \rho_{ee} \op{e}{e} ,
    \end{eqnarray}  
where $\cal{I}_\mathrm{Im}$ and $\cal{I}_\mathrm{Re}$ are those given by Eqs.~\eqref{integralfactors}. 



A model for an ensemble of solid-state defects, each coupled to its own vibrational mode, is given by taking multiple copies of the Hamiltonian in Eq.~(\ref{BertHam}). Additionally, the local modes may couple to one another through the long range acoustic phonons, which induces a set of nonlocal normal modes. 
This suggests that our results on structured environments in Sec.~\ref{sec:structured} on symmetric subspace protection and the existence of a privileged symmetric mode may be applicable. 
If lattice couplings between phononic modes are negative, the energy of the symmetric mode will be lower and gapped from the other modes. The higher energies of the non-symmetric modes are likely to lead to faster decay into the thermal bath, resulting in those modes decaying quickly to their steady state. 
The existence of such a gapped mode also allows the possibility of manipulating the spins via that mode.

The exact dynamics of such a system depends highly on the frequencies, the inter-mode coupling rates, and spin-mode coupling rates, which determine whether the system operates in the underdamped or overdamped regime (see Sec.~\ref{asymptotic}). 
For many solid-state defect emitters, such as NV centres, these frequencies and couplings are not well known (and, in some cases, are contradictory) and will differ depending on whether, for example, the system being considered is bulk or nanocrystalline. This suggests our model could be used to constrain these parameters. Our model predicts very different results for different phonon-phonon coupling rates, mode coupling strengths and oscillator values. For example, one could use our model in combination with experimental measurements to determine whether the defect system is in the underdamped or overdamped regime or to place bounds on the coupling and decay rates.

\section{Conclusion}

In this work, we have solved the spin-boson model for the case of a single large spin coupled to a collection of bosonic modes. We provide an analytic solution for the non-Markovian reduced spin dynamics that applies when the modes themselves can decay into local thermal environments. We identify two regimes of interest: underdamped and overdamped. In the underdamped regime, spin coherences oscillate while decaying at a rate proportional to the modes' decay rates. In the overdamped regime, the coherences experience no oscillations and decay at a rate \emph{inversely} proportional to the modes' decay rate. This Zeno-like effect can serve as a mechanism to preserve spin coherences. These regimes may also determine whether the dynamics is non-Markovian Markovian~\cite{Thoss2021}.


In the multiple spin-boson setting where the modes are intercoupled, the existence of normal modes with a fast decay into the thermal bath can result in protective effects on the \emph{collective} spin. For two spins, population transfer out of symmetric subspace depends only on the decay of the antisymmetric normal mode, and the subspace can be preserved for significantly longer than expected when this decay is large.
For $N$ coupled bosonic modes, equal coupling yields a single privileged symmetric normal mode whose energy is gapped from the other $N-1$ degenerate normal modes. 
Remarkably, this holds when the couplings between the modes are random in magnitude with a common mean and variance. In this case the energy gap persists between a single privileged near-symmetric normal mode and all the other modes whose energies are clustered within some energy window. 
This energy gap allows for the possibility of using dynamical decoupling to perform quantum information processing procedures, such as engineering effective spin squeezing or geometric phase interactions between the spins.

Finally, we connect our analysis to a simple model for defects in solid-state systems and discuss where our solutions and analyses can be applied. Some physical settings, such as NV centres in diamond, have extremely large vibrational decay rates, which could interfere with coherent effects. Our results suggest the opposite: large decay rates may actually serve to preserve inter-emitter coherences in a Zeno-type fashion. This gives a possible reason why recent optical experiments with NV centres in nanodiamonds have displayed collective effects~\cite{JuanBradBesg16, BradJohnBreu17} that require coherences on timescales much longer than that of the corresponding vibrational decay. Collective effects in the optical degrees of freedom are present in a wide range of solid-state systems, including rare-earth-ion doped solids, molecular aggregates, and low-dimensional excitonic solids \cite{cong_dicke_2016}, and it will be interesting to apply our findings in these contexts.

\acknowledgments
The authors thank Akib Karim for insightful discussions. B.Q.B., T.V., and G.K.B. acknowledge support from the Australian Research Council Centre of Excellence for Engineered Quantum Systems (Grant No.\ CE 170100009).
B.Q.B. was additionally supported by the Australian Research Council Centre of Excellence for Quantum Computation and Communication Technology (Project No.\ CE170100012) and the Japan Science and Technology Agency through the MEXT Quantum Leap Flagship Program (MEXT Q-LEAP).

\appendix

\begin{widetext}
\section{Removing time ordering using the Magnus expansion} \label{appendix:Magnus}
The Magnus expansion~\cite{MagnusExpansion2010} allows us to write a time-ordered exponential generated by a time-dependent operator $\hat{A}(t)$ in terms of a \emph{non}-time-ordered exponential,
    \begin{align} \label{Magusexpansion}
        {\cal{T}} \exp \left[\int_0^t dt' \hat{A}(t') \right] = \exp \big[ \hat\Omega(t) \big]   ,
    \end{align}
where $\hat\Omega(t)$ is a sum of terms related to the commutator of $\hat{A}(t)$ at different times,
    \begin{align} \label{Magnusterms}
        \hat\Omega(t) = \int_0^t dt_1 \, \hat{A}(t_1) - \frac{1}{2} \int_0^t dt_1 \, \int_0^{t_1} dt_2 \, [\hat{A}(t_2), \hat{A}(t_1) ] + \dots ,
    \end{align}
where additional terms involve nested multi-time commutators.
    
The unitary propagator in \erf{timeorderedpropagator} contains a time-ordered exponential operator over the vibrational modes. Since the modes are independent, $[\hat{v}_k, \hat{v}_{k'}^\dagger ] = \delta_{k,k'}$, 
we can treat each vibrational mode, indexed by $k$, separately. For each $k$, the time-ordered exponential is generated by 
$\hat{V}_k(t) \coloneqq \eta_k \hat{v}^{\dagger}_k e^{i \omega_k t} + \eta^*_k \hat{v}_k e^{-i \omega_k t}$,
which satisfies the following multi-time commutator,
        \begin{align}
        [\hat{V}_k(t_2), \hat{V}_k(t_1) ] 
        & = 2 |\eta_k|^2 \sin [\omega_k (t_1-t_2)]   .
    \end{align}
All higher-order commutators vanish, so the Magnus expansion in Eqs.~(\ref{Magusexpansion}--\ref{Magnusterms}) requires only first- and second-order terms.
Using this fact, we get for the time-ordered exponential,
\begin{align}
{\cal{T}} \exp \left[ -i m \int_0^t dt' \, \hat{V}(t') \right] 
&=
\prod_k {\cal{T}} \exp \Big[ -i m \int_0^t  dt'\, \hat{V}_k(t')  \Big]
\\
%
%
&= \exp \left[ - i \sum_k |\eta_k|^2 m^2  \int_0^t \! dt_1 \int_0^{t_1} \! dt_2 \, \sin [\omega_k (t_1 -t_2)] \right] \exp \left[ -i m \int_0^t dt'\,  \hat{V}(t')  \right]
\\
&= \exp \left[ - i m^2 \int d\omega \int_0^t \! dt_1 \int_0^{t_1} \! dt_2\,  J(\omega) \sin [\omega (t_1 -t_2)] \right] \exp \left[ -i m \int_0^t dt'\,  \hat{V}(t')  \right] .
\end{align}
In the second line, we converted the product of exponentials back into an exponential of a sum over terms. In the final line, we introduced an integral over $\omega$ with the spectral density $J(\omega)$, \erf{spectraldensity}, as the integration kernel. Importantly, note that all the time ordering of operators has been removed, with the effects captured in the exponential two-time integrals in the first term. In the main text and in the following Appendix, we express the final line as (\ref{propagatorv2}), where we have collected the terms in the first exponential into a $c$-number phase $\Phi(t)$, \erf{cnumberphase}.

\section{Trace over vibrational modes in a thermal state}\label{Appendix:thermaltrace}

When the vibrational modes are prepared in the mode-wise tensor-product thermal state in \erf{thermalstatetensorprod}, the partial trace in \erf{traceintegral} can be performed analytically. We follow here the calculation in Agarwal (2010)~\cite{Agarwal2010}. 
For this tensor-product state, the $\mathcal{S}(t)$ factor describing the partial trace that appears in the expression for the reduced-spin matrix elements factorizes,
    \begin{align} \label{appendixtraceintegral}
        \mathcal{S}(t) 
        & = \prod_k {\mathrm{Tr}}_{k} \bigg[ \exp \Big[ -i (m-m') \int_0^t dt'\, (\eta_k \hat{v}^{\dagger}_k e^{i \omega_k t'} + \eta^*_k \hat{v}_k e^{-i \omega_k t'})\Big] \hat \rho_{\text{therm},k} \bigg]  ,
    \end{align}
and each mode is initially described by a thermal state with average excitation $\bar{n}_k$, \erf{thermalstatek}.
We will treat each trace separately. Defining for convenience the coefficients
    \begin{align} \label{etakdef}
         \eta_k(t) \coloneqq -(m-m')\int_0^t dt'\, \eta_k  e^{i \omega_k t}  ,
    \end{align}
the exponential operator in the above expression can be written in the disentangled form,
    \begin{align} 
        \exp \big[ i \eta_k(t) \hat{v}^{\dagger}_k + i \eta_k^*(t) \hat{v}_k \big] 
        =
        \exp \big[ i \eta_k(t) \hat{v}^{\dagger}_k \big] \exp \big[ i \eta_k^*(t) \hat{v}_k \big] 
        \exp \big[ -\smallfrac{1}{2} |\eta_k(t)|^2 \big] .
    \end{align}
The trace can now be taken trivially,
 \begin{align} 
        (\text{trace over vib. mode $k$})
        & = \frac{1}{\pi \bar{n}_k} \exp \big[ -\smallfrac{1}{2} |\eta_k(t)|^2 \big]
        \int d^2\alpha \, \exp \left( -\frac{|\alpha|^2}{\bar{n}_k} \right) \exp \big[ i \eta_k(t) \alpha^* + i \eta_k^*(t) \alpha \big]
        \\
        & = 
         \exp \left[ -\big(\bar{n}_k + \tfrac{1}{2} \big)|\eta_k(t)|^2 \right]
        \\
        & = 
         \exp \left[ - \frac{1}{2} \coth \left( \frac{\hbar \beta \omega_k }{2} \right) (m-m')^2 \int_0^t dt_1\,  \int_0^t dt_2\, e^{i \omega_k (t_1 - t_2)} \right] ,
    \end{align}
where we substituted for $|\eta_k(t)|^2$ using \erf{etakdef} and rewrote the thermal factor using
    \begin{equation}
        \bar{n}_k + \frac{1}{2} = \frac{1}{2} \coth \left( \frac{\hbar \beta \omega_k }{2} \right) .
    \end{equation}

Summing this expression over all vibrational modes (in the exponential) and including the spectral density $J(\omega)$, \erf{spectraldensity}, we find the integral in \erf{appendixtraceintegral} to be
    \begin{align} \label{appendixSfinal}
        \mathcal{S}(t) 
          &= 
          \exp \bigg[ -i (m-m')^2
          \int_0^t dt_1\, \int_0^t dt_2\,\int d\omega \,
          J(\omega)
          \coth \left( \frac{\hbar \beta \omega }{2} \right)
          \cos[ \omega (t_1 - t_2)]
          \bigg] .
    \end{align}   
    
\section{Derivation of the correlation function} \label{Appendix:correlation}   
First, consider the quadrature correlation function for a single mode $k$ without coupling to a dissipative bath.
Then, due to the free evolution of the mode $\hat{v}_k(t)=e^{-i\omega_k t} \hat{v}_k(0)$ we have 
\begin{equation}
\expt{\hat{X}_k(t) \hat{X}_k(0)}
=
{\rm Tr}_{V}[(\eta_k e^{-i\omega_k t}\hat{v}_k(0)+\eta^*_k e^{i\omega_k t}\hat{v}^{\dagger}_k(0))(\eta_k \hat{v}_k(0)+\eta^*_k \hat{v}^{\dagger}_k(0))\hat{\rho}_V(0)]
\end{equation}

The initial state of the modes, $\hat{\rho}_V(0)$, is that of Eq.~\eqref{thermalstatetensorprod}, where each mode is thermally occupied with inverse temperature $\beta$ and mean occupation number $\bar{n}_k=1/(e^{\beta \omega_k}-1)$. The only surviving terms in the trace are those with equal numbers of creation and annihilation operators; hence,
\begin{align}
\expt{\hat{X}_k(t) \hat{X}_k(0)}
&=|\eta_k|^2 (e^{-i \omega_k t}{\rm Tr}_V[(\hat{v}^{\dagger}_k(0)\hat{v}_k(0)+1) \hat{\rho}_V(0)]+e^{i \omega_k t}{\rm Tr}_V[\hat{v}^{\dagger}_k(0)\hat{v}_k(0) \hat{\rho}_V(0)]\\
&=
|\eta_k|^2 \big[ e^{-i \omega_k t} (1+\bar{n}_k)+e^{i \omega_k t}\bar{n}_k \big] \\
&=|\eta_k|^2 \big[ \coth(\beta \omega_k/2) \cos(\omega_k t)-i\sin(\omega_k t) \big]. 
\end{align}

When dissipation is included,
the quadrature correlation function can be computed starting with the joint density matrix for mode $k$ and its local environment $E$, $\hat{\rho}_{V}(0)\otimes \hat{\rho}_{E}(0)$ (mode label $k$ suppressed) evolve according the total Hamiltonian $\hat{H}_{V E}$, which includes local and interaction couplings, and finally tracing:
\begin{align}
     \expt{\hat{X}_k(t) \hat{X}_k(0)}
     &= 
     {\rm Tr}_{VE}\Big( e^{i \hat{H}_{V E}t} \big[ \eta_k \hat{v}_k(0)+\eta^*_k \hat{v}^{\dagger}_k(0) \big] e^{-i \hat{H}_{V E}t} \big[ \eta_k \hat{v}_k(0)+\eta^*_k \hat{v}^{\dagger}_k(0) \big] 
     \hat{\rho}_{V}(0)\otimes \hat{\rho}_{E}(0) \Big)
     \\
     &=
     {\rm Tr}_{V} \Big\{ 
     {\rm Tr}_{E} \Big( e^{-i \hat{H}_{V E}t} \hat{\rho}_{E}(0) e^{i \hat{H}_{V E}t} \big[ \eta_k \hat{v}_k(0)+\eta^*_k \hat{v}^{\dagger}_k(0) \big] \Big)
     \big[ \eta_k \hat{v}_k(0)+\eta^*_k \hat{v}^{\dagger}_k(0) \big] 
     \hat{\rho}_V(0) \Big\}.
\end{align}

We consider both the modes and their environments to be in separable thermal states characterized by inverse temperature $\beta$ --- that is, thermal equilibrium.
When the environment satisfies the conditions that giving rise to Lindblad evolution of the mode, i.e. the evolution satisfies the Born-Markov approximation, the trace over $E$ gives rise to decay of the mode operators $\hat{v}_k(t)=e^{-i\omega_kt}e^{-\Gamma_k t/2} \hat{v}_k(0)$. 
Note that the bath temperature does not affect the time evolution of $\hat{v}_k$, although it does in general affect other moments (such as $\hat{n}_k$).
Assuming as before that the modes are thermal occupied, the correlation function becomes
\begin{align}
\expt{\hat{X}_k(t) \hat{X}_k(0)}&=e^{-\Gamma_k t/2}|\eta_k|^2 (e^{-i \omega_k t}{\rm Tr}_V \big[(\hat{v}^{\dagger}_k(0)\hat{v}_k(0)+1) \hat{\rho}_V(0)]+e^{i \omega_k t}{\rm Tr}_V[\hat{v}^{\dagger}_k(0)\hat{v}_k(0) \hat{\rho}_V(0) \big]
\\
&= e^{-\Gamma_k t/2}|\eta_k|^2 \big[ e^{-i \omega_k t} (1+\bar{n}_k)+e^{i \omega_k t}\bar{n}_k \big] 
\\
&= e^{-\Gamma_k t/2}|\eta_k|^2 \big[ \coth(\beta \omega_k/2) \cos(\omega_k t)-i\sin(\omega_k t) \big] . 
\end{align}

\section{Decay of the collective spin projectors} \label{AppendixProjs}
For multiple spin-$\frac{1}{2}$ particles, the coupled and local spin bases are related by Clebsch-Gordan coefficients. For two spins, the collective states $\ket{J,M}$ are given in terms of the local states $\ket{m_1,m_2}$ as 
    \begin{align}
        \ket{J=1,M=1} &= \ket{\tfrac{1}{2}, \tfrac{1}{2}} \\
        \ket{J=1,M=0} &= \tfrac{1}{\sqrt{2}} \left( \ket{\tfrac{1}{2}, -\tfrac{1}{2}} + \ket{\tfrac{1}{2}, -\tfrac{1}{2}} \right) \\
        \ket{J=1,M=-1} &= \ket{-\tfrac{1}{2}, -\tfrac{1}{2}} \\
        \ket{J=0,M=0} &= \tfrac{1}{\sqrt{2}} \left( \ket{\tfrac{1}{2}, -\tfrac{1}{2}} - \ket{\tfrac{1}{2}, -\tfrac{1}{2}} \right) 
        \, .
    \end{align}
From these relations, the evolution of the collective-basis projectors using equation \erf{eqRhoM1M2SymAndAntiSym} is,
    \begin{align}
        \hat{P}_{J,0}(t) & = \frac{1}{2} \Big( \ketbra{\half, -\half}{\half, -\half} + \ketbra{-\half, \half}{-\half, \half} \Big)
         +\frac{(-1)^{J+1}e^{-4 \mathcal{I}_\text{Re}(t;\omega_-, \Gamma_-) }}{2}  
        \Big( \ketbra{\half, -\half}{-\half, \half} + \ketbra{-\half, \half}{\half, -\half} \Big) \\
        & = \frac{1}{2} \big[ 1 - (-1)^{J+1}e^{-4 \mathcal{I}_\text{Re}(t;\omega_-, \Gamma_-) } \big] \hat{P}_{1,0} + \frac{1}{2} \big[ 1 + (-1)^{J+1}e^{-4 \mathcal{I}_\text{Re}(t;\omega_-, \Gamma_-) } \big] \hat{P}_{0,0}  
        \, .
    \end{align}

\section{Connection to a solid-state electronic-vibrational model}
\label{ConnectHams}

Consider the vibronic Hamiltonian, Eq.~\eqref{BertHam}, but with multiple vibrational modes coupled to the two-level electronic emitter:
    \begin{align} \label{multiBetzHam}
        \hat{H}  &= \Omega \hat{j}_z +
        \sum _k \omega_k ( \hat{v}^{\dagger}_k \hat{v}_k + \tfrac{1}{2} )
        + \op{e}{e} \otimes \sum _k (\eta_k \hat{v}^{\dagger}_k + \eta_k^* \hat{v}_k) \nonumber
    \\
    &= \Omega \hat{j}_z +
     \sum _k \omega_k ( \hat{v}^{\dagger}_k \hat{v}_k + \tfrac{1}{2} )
        +  \hat{j}_z \otimes \frac{1}{2} \sum _k (\eta_k \hat{v}^{\dagger}_k + \eta_k^* \hat{v}_k) + \frac{1}{2} \sum _k (\eta_k \hat{v}^{\dagger}_k + \eta_k^* \hat{v}_k)
    \end{align}
In the second line, we have used that fact that $\op{e}{e} = \frac{1}{2}\big(\hat{j}_z + \hat{I}\big)$ for spin $j=1/2$.
One identifies this Hamiltonian as nearly the same as our original spin-boson Hamiltonian with a few modifications. First, $\eta_k \rightarrow \eta_k/2$. More importantly, we find an additional coherent drive on the modes of strength $|\eta_k|$. However, we do not need to rely on this identification in order to solve for the reduced spin state, as we show below. 

Following the derivation in Sec.~\ref{secAnalyticDephasingResults}, the interaction-picture propagator using the Hamiltonian Eq.~\eqref{multiBetzHam} is
    \begin{equation}
        \hat{U}(t) = {\cal{T}} \exp \left[ -i \int _0 ^t dt'   \op{e}{e} \otimes \hat{V}(t') \right] ;
    \end{equation}
compare to the propagator in Eq.~\eqref{originalpropagator}. We use the solutions in the main text by identifying the eignenvalues $m = 1$ for $\op{e}{e}$ and $m=0$ for $\op{g}{g}$. With this, the formal solution for the joint electronic-vibrational state at time $t$ is found from Eq.~\eqref{jointstatesolution},
    \begin{align}
       \hat \rho(t) 
       &= \rho_{gg} \op{g}{g} \otimes \hat{\rho}_V(0) + \rho_{ge}  \otimes e^{ -i \Phi(t)}\op{g}{e} \hat \rho_V(0) e^{i \int_0^t dt'\, \hat{V}(t')} \nonumber \\
            & + \rho_{eg} \op{e}{g} \otimes e^{ i \Phi(t)} e^{ -i \int_0^t dt'\, \hat{V}(t')} \hat \rho_V(0) 
            + \rho_{ee} \op{e}{e} \otimes e^{ -i \int_0^t dt'\, \hat{V}(t')} \hat \rho_V(0) e^{i \int_0^t dt' \hat{V}(t')} 
        ,
    \end{align}
where $\rho_{jk} \coloneqq \bra{j} \hat{\rho}^\text{el} (0) \ket{k}$ are the matrix elements of the initial electronic state. 

Setting the initial state of the vibrational modes to be the multimode thermal state, Eq.~\eqref{thermalstatek}, then the reduced electronic state, $\hat \rho^\text{el}(t) \coloneqq \text{Tr}_\text{vib}[\hat \rho(t)]$, is found from Eq.~\eqref{eqRhoMMprimeSingleSpin} to be
    \begin{align}
       \hat \rho^\text{el}(t) 
       &= \rho_{gg} \op{g}{g} 
       + \rho_{ge} e^{  -i {\cal{I}}_{\mathrm{Im}}(t; \vec \omega) - {\cal{I}}_{\mathrm{Re}}(t; \vec \omega) } \op{g}{e} 
       + \rho_{eg} e^{  i {\cal{I}}_{\mathrm{Im}}(t; \vec \omega) - {\cal{I}}_{\mathrm{Re}}(t; \vec \omega) } \op{g}{e} 
       + \rho_{ee} \op{e}{e}  .
    \end{align}    
We have written out the full reduced spin state because of its compact form for spin-$\frac{1}{2}$ (as opposed to presenting the solution matrix-elementwise as for the general case in text). 
As expected, the ground and excited states do not evolve, while the coherences evolve in a non-Markovian fashion.

\subsection{Including optical decay and additional dephasing }
The Lindblad maps for a two-level system undergoing additional pure dephasing at rate $\Gamma_{{\rm dp}}$ and optical decay into the vacuum at a rate $\Gamma_\text{op}$ are
    \begin{align}
        \mathcal{D}_{\rm dp}[\hat{\rho}] 
        &\coloneqq \Gamma_\text{dp} \left( \hat{\sigma}_z \hat{\rho} \hat{\sigma}_z -  \hat{\rho} \right) , \\
        \mathcal{D}_\text{op}[ \hat{\rho}]
        &\coloneqq \Gamma_\text{op} \left( \hat{\sigma}_- \hat{\rho} \hat{\sigma}_+ - \tfrac{1}{2} \hat{\sigma}_+ \hat{\sigma}_- \hat{\rho} - \tfrac{1}{2} \hat{\rho} \hat{\sigma}_+\hat{\sigma}_- \right) 
        ,
    \end{align}
where $\hat{\sigma}_- \coloneqq \op{g}{e}$, $\hat{\sigma}_+- \coloneqq \op{e}{g}$, and $\hat{\sigma}_z \coloneqq \op{e}{e}-\op{g}{g}$.
The dephasing map simply generates decay of the coherences at rate $\Gamma_\text{dp}$.
The final two terms on the right-hand side of the optical decay map are integrated directly in our solution in the standard fashion as a anti-Hermitian Hamiltonian $\hat{H} = i \frac{\Gamma_\text{op} }{2} \op{e}{e}$.
The first term describes incoherent refeeding of the ground state $\ket{g}$ directly from the excited state $\ket{e}$. 
Both effects can be included directly into the solution above to obtain
    \begin{align} \label{opticaldecayequation}
       \hat \rho^\text{el}(t) 
       &= (1 - \rho_{ee} e^{-\Gamma_\text{op} t}) \op{g}{g} 
       + \rho_{ge} e^{-(\Gamma_\text{dp}+\frac{\Gamma_\text{op}}{2} )t} e^{  -i {\cal{I}}_{\mathrm{Im}}(t; \omega, \Gamma) - {\cal{I}}_{\mathrm{Re}}(t; \omega, \Gamma } \op{g}{e} \nonumber \\
       & \quad + \rho_{eg}e^{-(\Gamma_\text{dp}+\frac{\Gamma_\text{op}}{2}) t}  e^{  i {\cal{I}}_{\mathrm{Im}}(t; \omega, \Gamma) - {\cal{I}}_{\mathrm{Re}}(t; \omega, \Gamma) } \op{e}{g} 
       + e^{-\Gamma_\text{op} t} \rho_{ee} \op{e}{e} .
    \end{align}  
The apparent disappearance of the initial matrix element $\rho_{gg}$ comes from the fact that the $\op{g}{g}$ coefficient is $\rho_{gg} + \rho_{ee}(1 - e^{-\Gamma_\text{op} t}) = 1 - \rho_{ee} e^{-\Gamma_\text{op} t}$, where we used $\rho_{gg} + \rho_{ee} = 1$.
The above can be compared to the reduced-state equation derived by Betzholz \emph{et al.}~\cite{BetzTorrBien14}, noting that their solution is in the Schr\"{o}dinger picture, while Eq.~\eqref{opticaldecayequation} is in the interaction picture with respect to the bare electronic and vibrational Hamiltonians. We also note $\tilde{\Gamma}$ should be $\tilde{\Gamma}/2$ in Eq.~(75) of \cite{BetzTorrBien14}.

\end{widetext}

\bibliography{NVcollective_refs.bib}



\end{document}